# Radioisotopes production using lasers: from basic science to applications.

**M. R. D. Rodrigues[1], A. Bonasera[1,2*], M. Scisciò[3], J. A. Pérez-Hernández[4], M. Ehret[4], F. Filippi[3], P. L. Andreoli[3], M. Huault[5], H. Larreur[5,6,7], D. Singappuli[6], D. Molloy[7,8], D. Raffestin[6], M. Alonzo[3], G. G. Rapisarda[2,9], D. Lattuada[2,10], G. L. Guardo[2], C. Verona[11], Fe. Consoli[2], G. Petringa[2], A. McNamee[8], M. La Cognata[2], S. Palmerini[12,13], T. Carriere[6], M. Cipriani[3], G. Di Giorgio[3], G. Cristofari[3], R. De Angelis[3], G. A. P. Cirrone[2], D. Margarone[8,14], L. Giuffrida[2,14], D. Batani[6], P. Nicolai[6], K. Batani[15], R. Lera[4], L. Volpe[4,16], D. Giulietti[17], S. Agarwal[18], M. Krupka[18,19], S. Singh[18,19], Fa. Consoli[3**]**

[1] Cyclotron Institute, Texas A&M University, College Station, TX , USA

[2] Laboratori Nazionali del Sud, Istituto Nazionale di Fisica Nucleare (LNS-INFN), Catania, Italy

[3] ENEA, Fusion and Technologies for Nuclear Safety and Security Department-C, Frascati, Italy

[4] CLPU (Centro de Láseres Pulsados), Villamayor, Spain

[5] Universidad de Salamanca, Salamanca, Spain

[6] Université de Bordeaux, CNRS, CEA, CELIA (Centre Lasers Intenses et Applications), Talence, France

[7] HB11 Energy Holdings Pty, Freshwater, NSW, Australia

[8] Queen's University Belfast, School of Mathematics and Physics, Belfast, UK

[9] Dipartimento di Fisica e Astronomia "E. Majorana", Università di Catania, Catania, Italy.

[10] Facoltà di Ingegneria e Architettura, Università degli Studi di Enna "Kore", Enna, Italy

[11] Dipartimento di Ingegneria Industriale, Università di Roma "Tor Vergata", Roma, Italy

[12] Dipartimento di Fisica e Geologia, Università degli Studi di Perugia, Perugia, Italy

[13] Istituto Nazionale di Fisica Nucleare, sezione di Perugia, Perugia, Italy

[14] ELI Beamlines Facility, The Extreme Light Infrastructure ERIC, Dolni Brezany, Czech Republic

[15] IPPLM Institute of Plasma Physics and Laser Microfusion, Warsaw, Poland

[16] ETSIA, Universidad Politécnica de Madrid, Madrid, Spain

[17] Dipartimento Fisica, "E. Fermi", Università di Pisa and INFN, Pisa, Italy

[18] FZU-Institute of Physics of Czech Academy of Sciences, Prague, Czech Republic

[19] Institute of Plasma Physics of Czech Academy of Sciences, Prague, Czech Republic

**Correspondence:**
***** A. Bonasera
abonasera@comp.tamu.edu
****** Fa. Consoli
fabrizio.consoli@enea.it

**Abstract**

Laser technologies improved after the understanding of the Chirped Pulse Amplification (CPA) which allows energetic laser beams to be compressed to tens of femtosecond (fs) pulse durations and focused to few μm. Protons of tens of MeV can be accelerated using for instance the Target Normal Sheath Acceleration (TNSA) method and focused on secondary targets. In such conditions, nuclear reactions can occur and radioisotopes relevant for medical purposes be produced. High repetition lasers can be used to produce enough isotopes for medical applications. This route is competitive to conventional methods mostly based on accelerators. In this paper we study the production of $^{67}$Cu, $^{63}$Zn, $^{18}$F and $^{11}$C currently used in positron emission tomography (PET) and other applications. At the same time, we study the reaction $^{10}$B(p,α)$^{7}$Be and $^{70}$Zn(p,4n)$^{67}$Ga to put further constraints to the proton distributions at different angles and to the reaction $^{11}$B(p,α)$^{8}$Be relevant for energy production. The experiment was performed at the 1 petawatt (PW) laser facility at Vega III located in Salamanca-Spain. Angular distributions of radioisotopes in the forward (with respect to the laser direction) and backward directions were measured using a High Purity Germanium Detector (HPGE). Our results are reasonably reproduced by the numerical estimates following the approach of Kimura et al. (NIMA637(2011)167).

## 1 Introduction

Laser accelerated proton beams have been applied to radioisotopes production for medical purposes by several groups [1–4]. Kimura et al. analyzed in ref. [5] the feasibility of the laser production as an alternative procedure to other methods. Since then, the laser technology has largely improved and high repetition lasers in the petawatt regime are available at a cost competitive to accelerator-based systems. Those systems rely on the Chirped Pulse Amplification (CPA) method [6] pushing laser intensities to go beyond $10^{18}$ W/cm$^2$. At such intensities, ions from the target surface irradiated by the laser beams are accelerated to several MeV energies. This is due to the laser action which strips the less bound electrons from the target thus creating a strong (repulsive) electric field which accelerates the ions. Without going into the details of the mechanism, we will loosely refer to it as TNSA [1-5], we discuss an intuitive scenario which we put to an experimental test in this work. We assume that the stripped electrons leave a positively charged target. Ions or protons on the opposite sides of the surface of the target feel the repulsive Coulomb field, get accelerated and leave the target area thus decreasing the Coulomb repulsion. In particular, the shorter is the laser pulse, i.e., the faster the laser energy is released on the target, the quicker the positive electric field at the target increases. Thus, we may expect higher protons energies for shorter pulses and thin targets [4]. The accelerated ions will move in the direction perpendicular to the target and in *both* directions (maybe with different intensities), i.e., in the forward and backward direction with respect to the laser one. With protons reaching up to 60 MeV and above [7] isotope production becomes possible, and we will discuss experimental results on the following systems:

(a) $^{63}$Cu(p,n)$^{63}$Zn, Q = -4.15 MeV;
(b) $^{70}$Zn(p,4n)$^{67}$Ga, Q = -27.68 MeV;
(c) $^{70}$Zn(p,α)$^{67}$Cu, Q = 2.62 MeV;
(d) $^{10}$B(p,α)$^{7}$Be, Q = 1.14 MeV;
(e) $^{11}$B(p,n)$^{11}$C, Q = -2.76 MeV;
(f) $^{18}$O(p,n)$^{18}$F, Q = -2.44 MeV.

It is important to notice that the reactions above have positive and negative Q values, thus from their measurement we will test the proton distribution at different energies [5]. For instance, reaction (b) which has a relatively high threshold energy, requires proton energies above 28 MeV to be produced, while reaction (d) with a positive Q value could occur at any proton energy in principle. We would like to stress that in this work we will mainly concentrate on the reactions (a-f). Another important study of the present experiment is the reaction $^{11}B(p,\alpha)^{8}Be$ (Q = 8.7 MeV) which will be discussed in detail elsewhere [8].

## 2 Experiment and data analysis

The experiment was performed at the Centro de Laseres Pulsados (CLPU)-VEGA [9] in Salamanca-Spain in November 2022. The laser beam energy was kept near the nominal maximum value of 30 J while the pulse duration was increased from the lowest possible value of 30 fs to 200 fs. A scheme of the scattering chamber is given in the Appendix A. Longer pulses were chosen to try to increase the proton yield at energies around 1 MeV where the cross section for the fusion reaction $^{11}B(p,\alpha)^{8}Be$ is highest and to improve the shot-to-shot performance stability. The beam was focalized to 6 μm on a 6 μm thick Al target to produce TNSA (or other mechanisms) protons coming from impurities in the target [4]. The facility can deliver 1 shot per second but the frequency was limited by the time to change the position on the Al target and/or other requirements from the experimental campaign [8]. Many shots were dedicated to determining the proton and other ions energy distributions and we took advantage of this possibility to activate different targets located at different angles. While fusion reactions and proton distributions will be discussed in detail elsewhere [8], we were interested in nuclear activations to test the theoretical prediction of ref. [5]. A positive comparison with ref. [5] would actually confirm that lasers are competitive with other accelerator-based methods to produce radioisotopes for medical, veterinary, and other applications.

Following refs. [5,10] we write the proton distribution function as:

$$\frac{d^2N_p}{dEd\Omega} = b(\Omega)\left(\frac{N_1}{\sqrt{\pi E T_1}} e^{-E/T_1} + \frac{N_2}{\sqrt{\pi E T_2}} e^{-E/T_2}\right) \qquad (1)$$

$N_1 = 1.2\ 10^{13}$, $T_1 = 3.3$ MeV, $N_2 = 2.3\ 10^{12}$, $T_2 = 13.5$ MeV are parameters fitted to the experimental results of ref. [10]. Notice that eq. (1) is a phenomenological parametrization of the data, and we are not suggesting that the acceleration is a thermal process. The parameter b(Ω) is added to reproduce the angular distributions of the reactions. When integrated over the solid angle we assume $\int b(\Omega)d\Omega = 1/5$, where 1/5 is the ratio of the laser energy at Vega III (30J) and in ref. [10] (150J). There may be a scaling due to the different pulse duration [11] but we will take care of any correction at different angles through b(Ω). Integrating eq.(1) in energy and angle with the above assumption gives the total number of protons, say up to 40 MeV maximum energy: $N_p = 1.3\ 10^{13}$, decreasing the higher energy limit to 20 MeV gives a difference of less than 1%. Considering a pulse duration of 200 fs we get a proton current $I_p \sim 10^7$ A which is an enormous current. If a secondary target is located very close to the Al target, as will be discussed below for some cases, then a plasma may be created by the proton flow. Furthermore, if we accept our naïve picture of the "TNSA mechanism" that electrons are ejected from the Al target in times of the order of the laser pulse duration, then those electrons reach the secondary target before the protons because of the much lighter mass. In this scenario the secondary target may become negatively charged in a short time and this may accelerate the positive ions further. The nuclear cross sections and ranges for different reactions maybe modified in such an environment and this could be the subject of future detailed investigations. Following the same steps above we can multiply eq. (1) by the proton energy and integrate over angle and energy. This gives the total energy provided to the protons $\langle E_p \rangle = 6.7(5.8)10^{12} MeV$ when integrating up to 40(20) MeV maximum proton energy. This result gives a conversion efficiency from laser energy to proton energy of about 3.1% *assuming* that an equal number of protons with the same energy distribution is produced in the back direction. The last assumption will be experimentally tested in this work. The average energy of the protons is $\frac{\langle E_p \rangle}{N_p} = 0.515\ MeV$ which would be almost optimal for the 3α fusion reaction. Of course, for reactions having a negative Q value the lower limit of integration of the proton distribution is given by -Q itself and this will cut all the low energy part of the proton distribution.

We can compare our 'benchmark' proton distribution function [10] with other data available in the literature. In Fig. 1, we compare to some selected cases obtained at the SULF laboratory [7] in China using a 2 PW laser focalized to about 6 μm like our case but at much higher power. The distribution function depends strongly on the detection angle with respect to the Al (or Cu) normal target decreasing of more than a factor 3 at 15 degrees. Furthermore, the proton high energy cutoff decreases when increasing the thickness from 4 to 10 μm. The distribution function from eq. (1) reproduces rather well the measured one in the region of interest for our work between 5 and 20 MeV. The fitted parameter, Fig. 1, maybe due to the combination of the laser energy and the pulse duration which are different in refs. [10,11]. We expect a similar angular and energy dependence in the back direction as well. A better comparison [8] could be done when the experimental data is measured over different angles and integrated over angles since the distribution function of refs. [5,10] is integrated over solid angle.

To calculate the number of reactions in the target is a more complicated process since we need to consider the proton energy distribution, the reaction cross section and the range of protons in the (assumed cold) target in the limit of thick targets. We can estimate this as:

$$\frac{dN}{d\Omega} = \rho \int_{E_{th}}^{E_{max}} \frac{dN_p}{dE d\Omega} \sigma(E) R(E) (1 - e^{-\frac{d}{R(E)}}) dE \qquad (2)$$

Where ρ is the (secondary) target density, d its thickness, $\sigma(E)$ is the reaction cross section of the process (taken from https://www.nndc.bnl.gov), R(E) is the range obtained from the SRIM (Stopping and Range of Ions in Matter software) [12], $E_{th}$ is the threshold energy for the reaction to occur (equal to zero for positive Q value reactions). The term in parenthesis inside the integral gives the probability that a proton collides with one nucleus of the target. When d<<R(E) we get:

$$\frac{dN}{d\Omega} = \rho d \int_{E_{th}}^{E_{max}} \frac{dN_p}{dE d\Omega} \sigma(E) dE \qquad (3)$$

In the limiting case that the proton distribution is a delta function of the energy (such as for accelerators), eq. (3) reduces to the familiar form used in nuclear physics applications to measure the cross section for instance [13]. Following the definition of the astrophysical S-factor, we parametrize the cross sections as:

$$\sigma(E) = \frac{S(E)}{E} e^{-\frac{b}{\sqrt{E}}} \qquad (4).$$

S(E) is written in terms of a Taylor expansion in energy [14]; for negative Q values reactions we write it as:

$$S(E) = \sum_i a_i (E + Q)^i ; S(E \to -Q) = 0. \qquad (5).$$

The free parameters $a_i$ and b are fitted to available experimental data, the number of parameters is chosen to have a good data reproduction. The parameterization in eq. (4) & (5) is especially important in energy regions where there is no data.

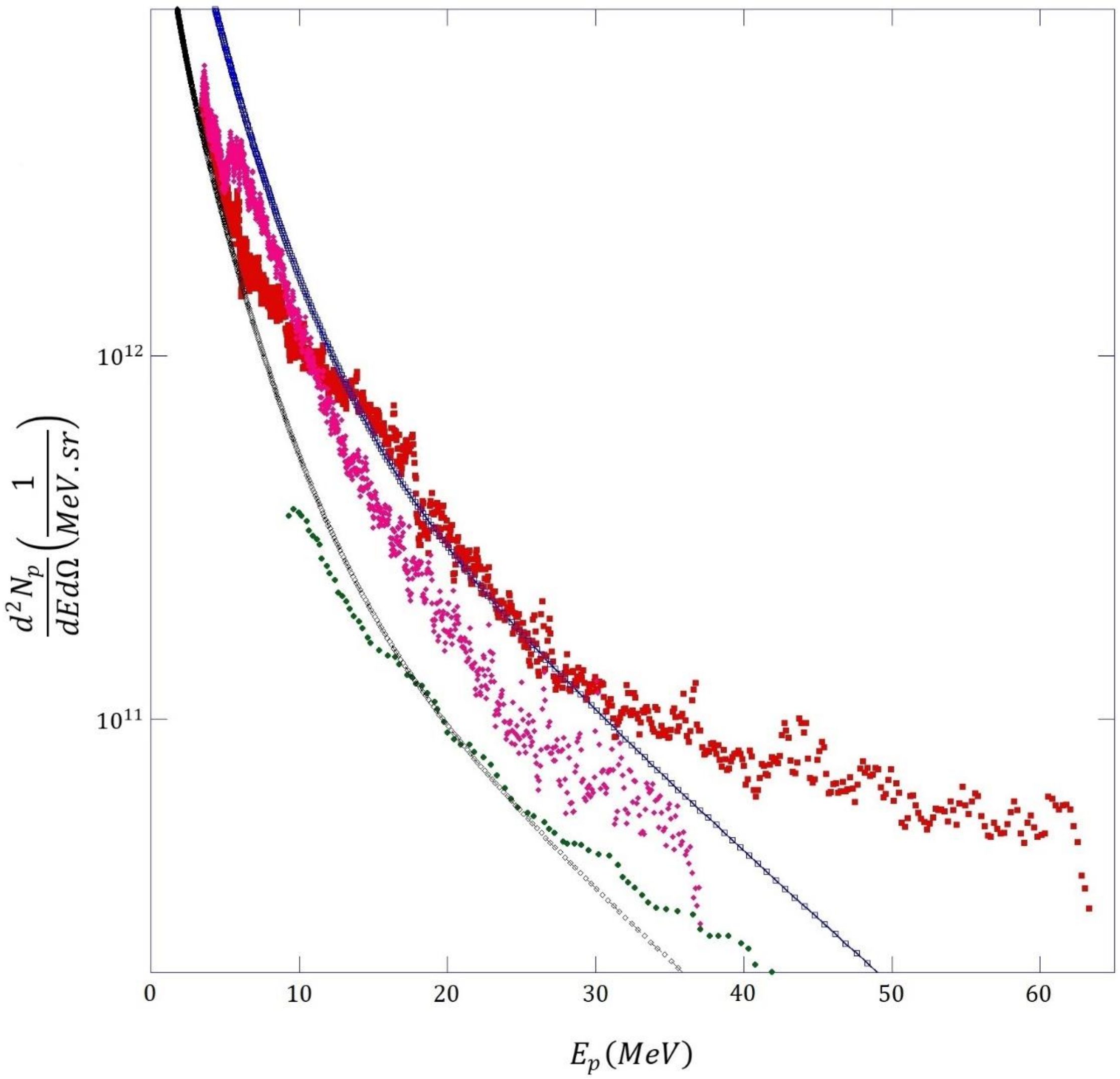

**Figure 1**. Proton distribution function (full symbols) from ref. [7]. The full (red) squares were obtained at zero degrees (with respect to the target normal) and the full (green) circles in the laser direction at 15°, the target thickness was 4μm. The full (pink) diamonds were measured at zero degrees but for a thickness of 10μm. Equation (1) was fitted to the 4μm cases with b(Ω)= 15(5) $sr^{-1}$.

In Table I, we will compare the results of eq. (2) integrated over angle to different reactions and $E_{max}$ = 40 MeV. These results are of guidance for the present experiment and maybe for future experiments as well.

| # | Reaction | Target | Thickness (mm) | N ($10^8$) | Q (MeV) | $E_\gamma$ (keV) | $T_{1/2}$ |
|---|---|---|---|---|---|---|---|
| 1 | $^{63}Cu(p,n)^{63}Zn$ | $^{nat}Cu$ | 1. | 1.4 | -4.15 | 511 | 38.5 m |
| 2 | $^{63}Cu(p,n)^{63}Zn$ | $^{63}Cu$ | 0.011 | 0.11 | -4.15 | 511 | 38.5 m |
| 3 | $^{70}Zn(p,4n)^{67}Ga$ | $^{70}Zn$ | 0.032 | $1.8\ 10^{-3}$ | -27.68 | 93.3 | 3.26 d |
| 4 | $^{70}Zn(p,\alpha)^{67}Cu$ | $^{70}Zn$ | 0.032 | $2.6\ 10^{-3}$ | 2.62 | 185 | 2.57 d |
| 5 | $^{10}B(p,\alpha)^{7}Be$ | $^{nat}B$ | 2. | 0.28 | 1.15 | 477.6 | 53.2 d |
| 6 | $^{11}B(p,n)^{11}C$ | $^{nat}B$ | 2. | 1.4 | -2.76 | 511 | 20.4 m |
| 7 | $^{11}B(p,\alpha)^{8}Be$ | $^{nat}B$ | 2. | 2.2 | 8.6 | - | $8\ 10^{-17}$ s |
| 8 | $^{12}C(p,X)^{11}C$ | CR39 | 1. | 0.1 | -16.5 | 511 | 20.4 m |
| 9 | $^{12}C(p,p)3\alpha$ | CR39 | 1 | 0.88 | -7.27 | - | - |
| 10 | $^{63}Cu(\alpha,X)^{65}Zn$ | $^{nat}Cu$ | 1 | - | -10.4 | 1115.5 | 244. d |
| 11 | $^{65}Cu(\alpha,X)^{68}Ga$ | $^{nat}Cu$ | 1 | - | -5.8 | 511 | 67.7 m |
| 12 | $^{18}O(p,n)^{18}F$ | $^{nat}B$ ($^{18}O$ $8.7\ 10^{-5}$ %) | 2. | 3 | -2.44 | 511 | 109.7 m |

**Table I.** Expected reaction rates N (corrected by concentration) using eq. (2) and assuming $\int b(\Omega)d\Omega = 1/5$. The $E_\gamma$ column refers to the highest intensity $\gamma$–ray emitted by the isotope; $T_{1/2}$ is the half-life [17].

There are some important features to notice.

i) The experimental production of $^{67}Ga$ indicates that there are protons of energy higher than 28 MeV (#3 of Table I).

ii) If there are enough $\alpha$-particles produced through the fusion reactions #5,7 in Table I, these could produce other reactions (ternary reactions) when impinging on other targets, for instance reactions #10,11 in Table I.

In fact, $\alpha$ particles of energy larger than 10 MeV can produce $^{65}Zn$ and this would indicate that their energy distribution is modified [18] by high energy protons [8]. Thus, to measure the distribution function of $\alpha$-particles is crucial and for this purpose, CR39 (Columbia Resin #39, a solid-state nuclear track detector) is mostly used since other methods are not effective due to the large number of protons and other particles produced in the plasma [8,15,16,18,19]. For this reason, we have estimated possible reactions occurring in the CR39 detector (which is rich in C ions) giving the same products coming from the main reactions such as 3$\alpha$, #9, and $^{11}$C, #8. As we can see from Table I those reactions are quite competitive especially if high energy protons are produced through the "TNSA mechanism". It is important to stress that reactions with negative Q values are often competitive with positive Q value ones. This is strictly dependent on the proton energy distribution (that can reach high values), the corresponding cross section and the range. In the following sections we will discuss in detail the main experimental findings and compare them to Table I.

## 2.1 $^{63}$Cu(p,n)$^{63}$Zn, $^{63}$Cu($\alpha$,X)$^{65}$Zn, $^{65}$Cu($\alpha$,X)$^{68}$Ga

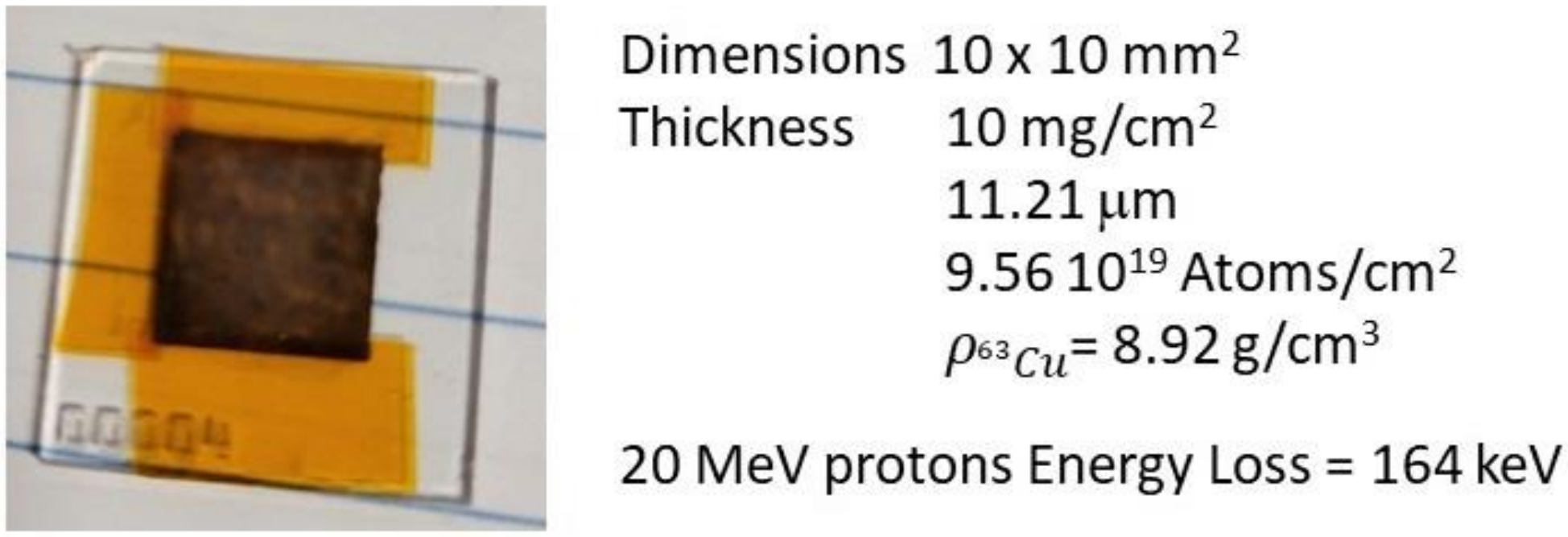

**Figure 2.** A picture and properties of the $^{63}$Cu target used in the experiment. The target is mounted on CR39 acting as a support and proton detector with pieces of Kapton® tape.

Thin $^{63}$Cu targets were mounted on CR39 both for support and to measure the energy distribution of the incident protons (and other ions if possible) which go through the target, see Fig. 2. The targets were located at different angles and distance with respect to the Al target. Notice that if high energy protons collide with the CR39, $^{11}$C maybe produced, reaction #8 in Table I, and this $\gamma$-decays with energy 511 keV, the same as the decay of $^{63}$Zn. In such cases the half-lives can be determined (or other signature $\gamma$s) when it is possible to distinguish the two cases.

It is also important to determine the number of reactions produced per shot since we were in multi-shot accumulation regime. Even if there may be large fluctuations from shot to shot [8], we can define an average over many shots $N_{av}$. The produced isotopes decay with the decay constant $\lambda$, thus we give a weight to each shot according to $\Pi = e^{-\lambda(t_0 - t_{shot})}$, and sum over all shots. $t_0$ is the starting measurement time using the HPGE detector and $t_{shot}$ is the time when the shot occurred, see Fig. 3, for more details see Appendix B.

In Table I, the reactions #1 and 2, it is clearly demonstrated that for thin targets the produced yield is very small (and comparable to the reaction rates obtained for CR39), thus we have repeated the analysis using a thick target of $^{nat}$Cu at different angles, see Fig. 4 for the case at zero degrees where the largest number of produced protons occurs. We expect an increase in the yield of at least an order of magnitude, Table I, and the higher statistics allow us a more detailed analysis.

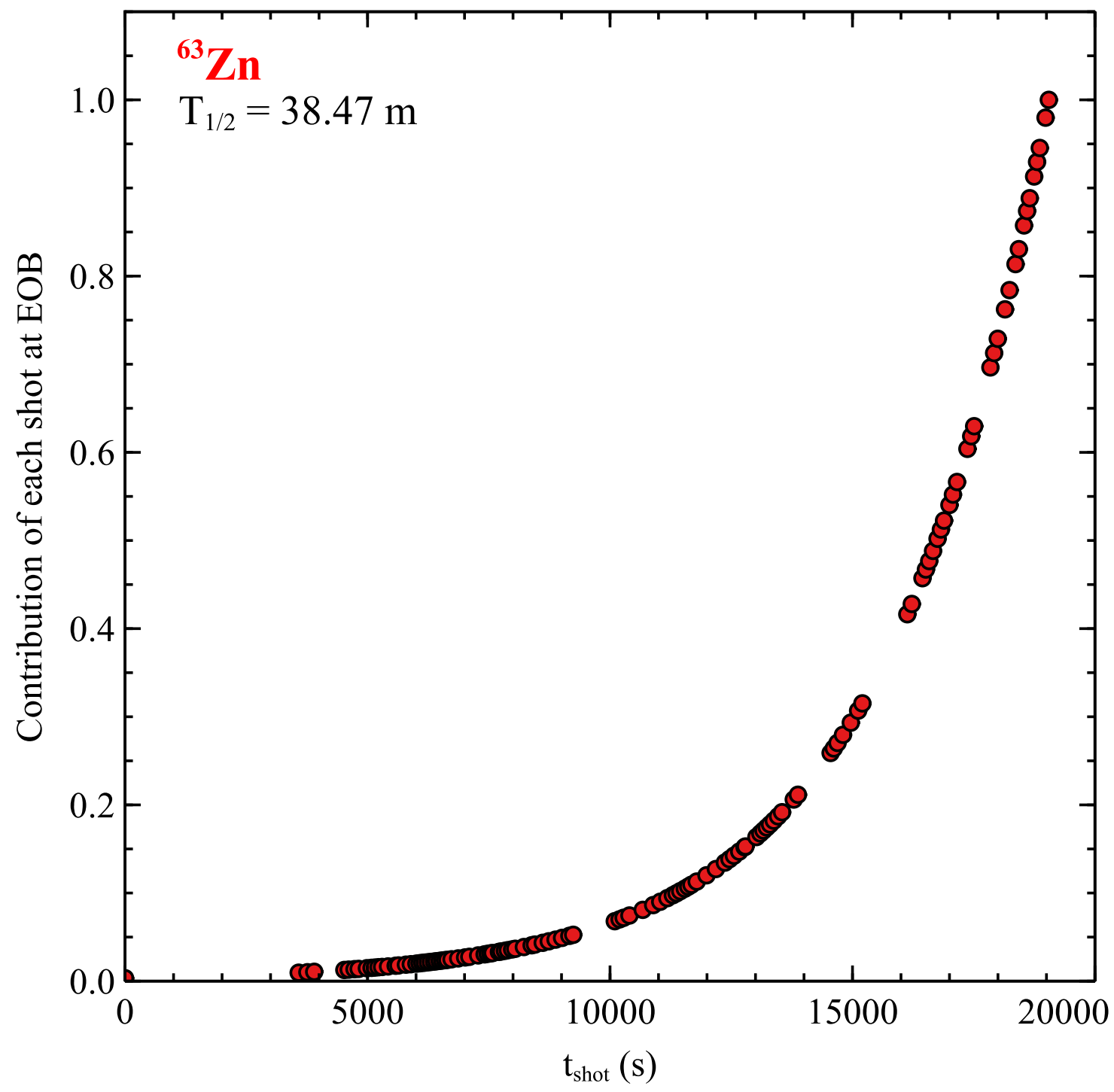

**Figure 3.** Probability contribution of each shot to the reaction $^{63}$Cu(p,n)$^{63}$Zn at End of Bombardment (EOB) as a function of shooting time starting from the first one ($t_{shot}$ = 0 s) to the last one when the chamber is opened and the sample decays were measured using the HPGE detector. The well-known half-life ($T_{1/2}$) of $^{63}$Zn is used to produce this result.

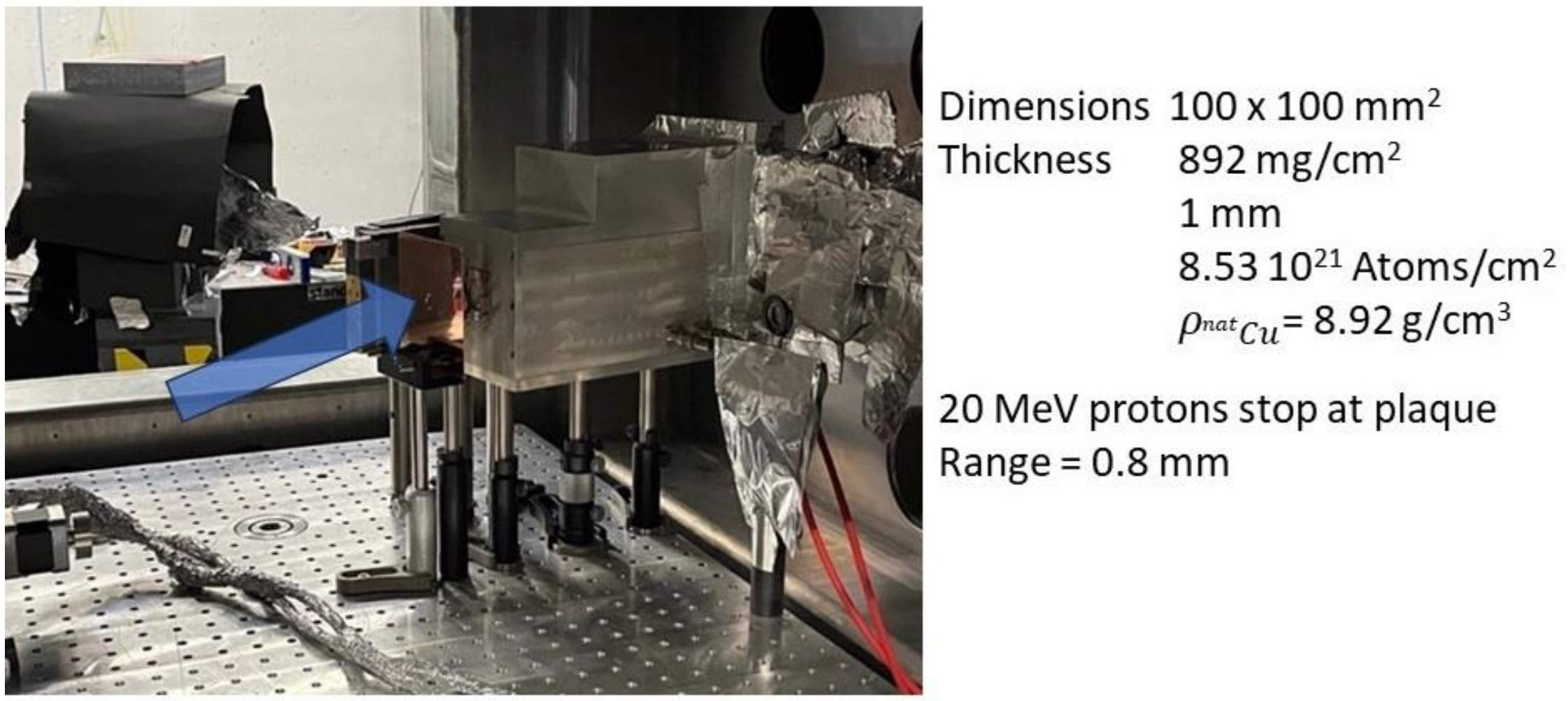

**Figure 4.** The $^{nat}$Cu target mounted in front of the Thomson parabola spectrometer. Notice the hole at the center of the target to let the energetic ions into the spectrometer. This position defines our zero degree where the largest number of protons, indicated in the figure by the arrow, is found. A diagram of the experiment can be found in Appendix A.

In Fig. 5, we show the background-subtracted β-delayed γ -ray spectrum using the thick $^{nat}$Cu target located at zero degrees with respect to the Al target perpendicular direction. At this angle, see Fig. 4, we expect the proton TNSA yield to be the highest. The full-energy-peak (FEP) at 511 keV is more intensely populated, a signature of $^{63}$Zn being produced, see Table I#1. Furthermore, other FEPs associated with lower intensity $^{63}$Zn decays are clearly observed.

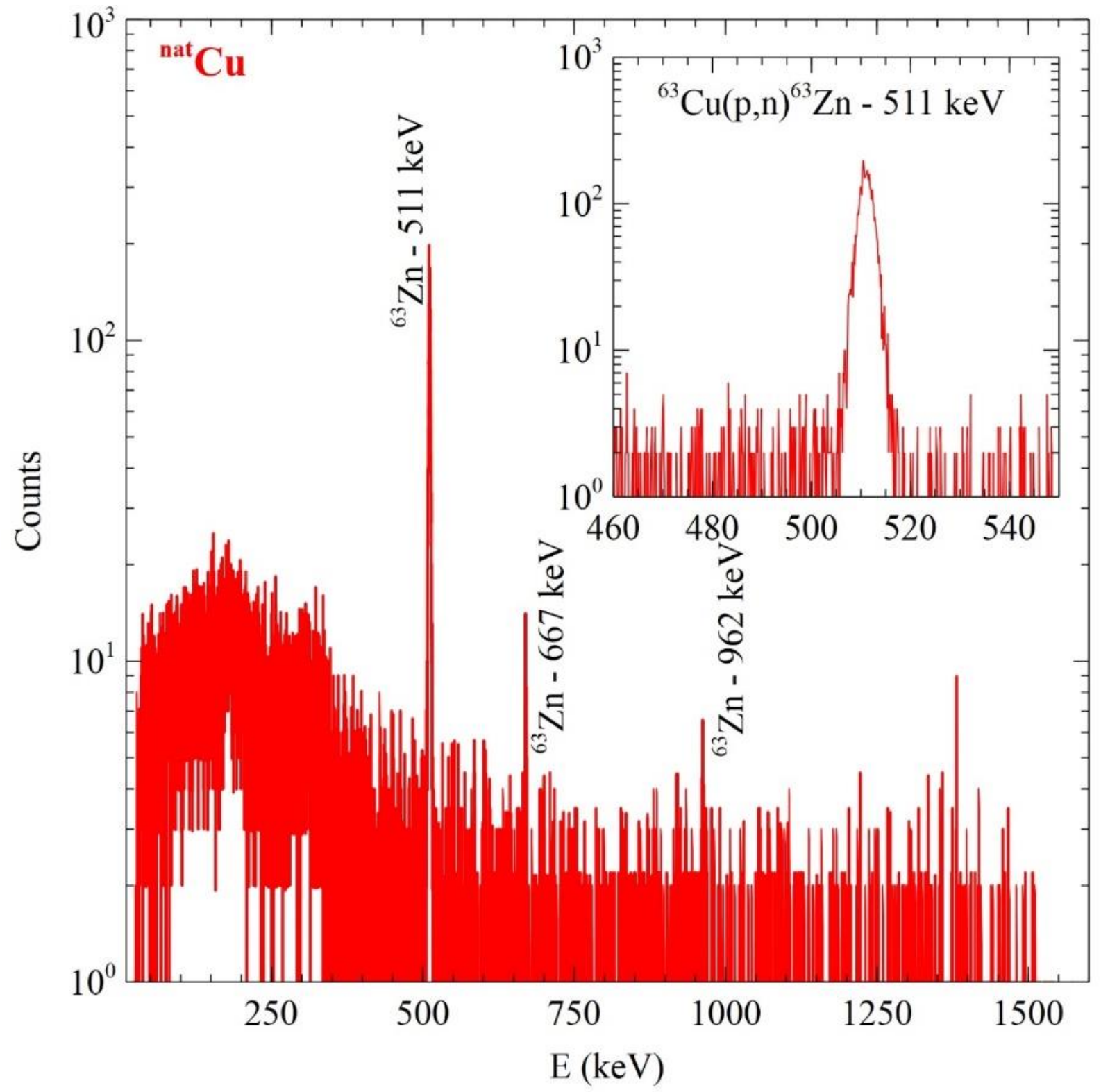

**Figure 5.** Background-subtracted β-delayed γ -ray spectrum using the thick $^{nat}$Cu target. The main contribution for the FEP at 511 keV, magnified in the inset, comes from the $^{63}$Zn decay produced in the $^{63}$Cu(p,n)$^{63}$Zn reaction. Other lower intensity γ associated to the $^{63}$Zn decay are also visible.

In Fig. 6, we plot the counting rate vs time for the zero-degree case where the statistics is large enough. We clearly see that the decay is in perfect agreement with the well-known half-life of $^{63}$Zn and demonstrate that there are no other competing reactions (for instance reactions with the $^{65}$Cu in the target).

Standard analysis techniques were applied to derive the number of produced isotopes at the end of irradiation, see for example refs. [20-23]. The final experimental uncertainties are simple propagation of the peak statistics, detector efficiencies and the fit uncertainties. The corrections due to the detector dead time, the backscattering peak and the respective summed coincidences, source self-absorption and isotopic enrichment, were also considered in the results, see Appendix B for more details.

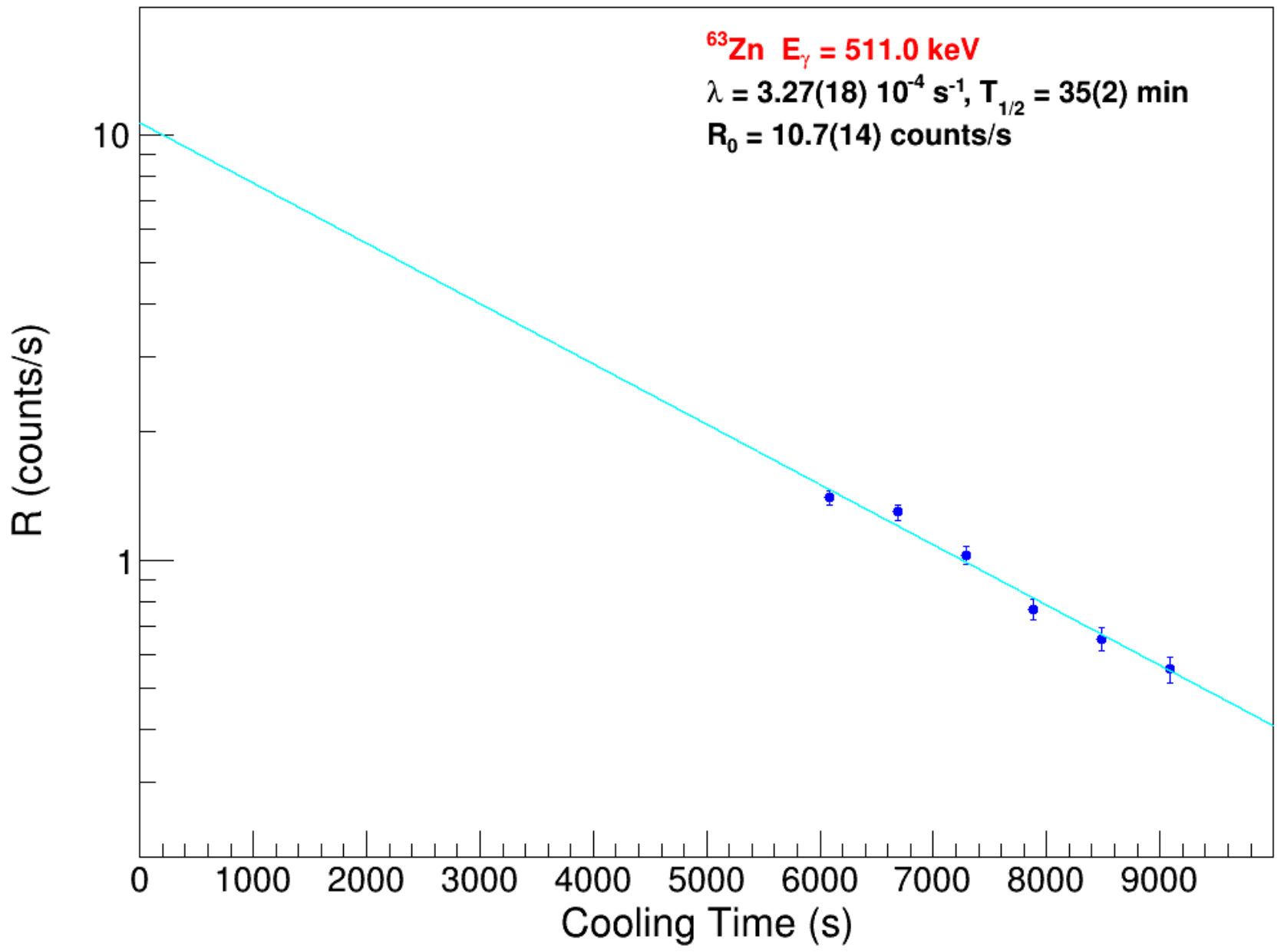

**Figure 6.** Counting rate decay as a function of Cooling Time. The decay constant (λ) fitted and the associated half-life ($T_{1/2}$) displayed in the figure confirms that the decaying nucleus is $^{63}$Zn. The fitted half-life $T_{1/2}$ is presented in the figure.

In Fig. 7, we plot the results for the Cu targets, full circles and diamonds. The $^{nat}$Cu gives a peak at 0 degrees as expected, while the thin $^{63}$Cu displays the largest value at 230 degrees. For the latter case we would expect a maximum at 180 degrees, but we could not install a target at such angle because of impeding optics. In any case, these results clearly demonstrate that the proton yield in the back direction is as strong as in the forward direction. Since the distribution is so strongly peaked, we cannot estimate the total yield integrated over the forward direction because we measured few angles. However, as we will show in a later section, we can estimate the opening angle of the produced protons in the forward direction [7,8,19] and obtain ΔΩ = 0.59±0.18 sr. This gives an average number of produced ions per shot N($^{63}$Zn) $\approx \frac{dN}{d\Omega}\Delta\Omega = (1.1 \pm 0.3)10^8$ in fair agreement with the estimate in Table I. Shooting at 1Hz for two hours at Vega III, we may get 345±94 MBq. Alternatively, one could use a lower energy laser but with higher repetition rate. For instance, Vega II in Salamanca-Spain, delivers 6J in 30fs at 10Hz, see also [19]. Assuming the same laser energy scaling as for Table I, this laser will produce twice as much radioisotope than the estimate above but at a much lower cost. A major problem in this approach may be connected to the time to align the "TNSA target" of choice at high frequency.

From the reaction #2 in Table I, we would expect the result at 230 degrees to be one order of magnitude smaller with respect to the zero degree because of the different targets thicknesses. Since the $^{63}$Cu target was mounted on (thick) CR39, see Fig. 2, we cannot exclude the contribution from $^{11}$C decays from reaction #8-Table I, if protons with energies above 16.5 MeV are produced at such angles. Unfortunately, for this case the statistics were poor and did not allow us to do a time evolution study as in Fig. 6. Other γ produced in the decay of $^{63}$Zn are comparable to the background at variance with the results of Fig. 5. Another reason could be that more protons are produced in the back direction [19] thus giving a higher yield than predicted. A detailed angular and energy distribution of the protons is not the goal of this work and will be discussed elsewhere [8]. The use of CR39 as a support for our thin targets will provide important constraints to the proton distribution for these angles. Furthermore, the choice of using a stack of two CR39, not clearly visible in Fig. 2, will allow to measure unambiguously high energy protons (if any) which went through the first CR39 and stopped in the second one [8].

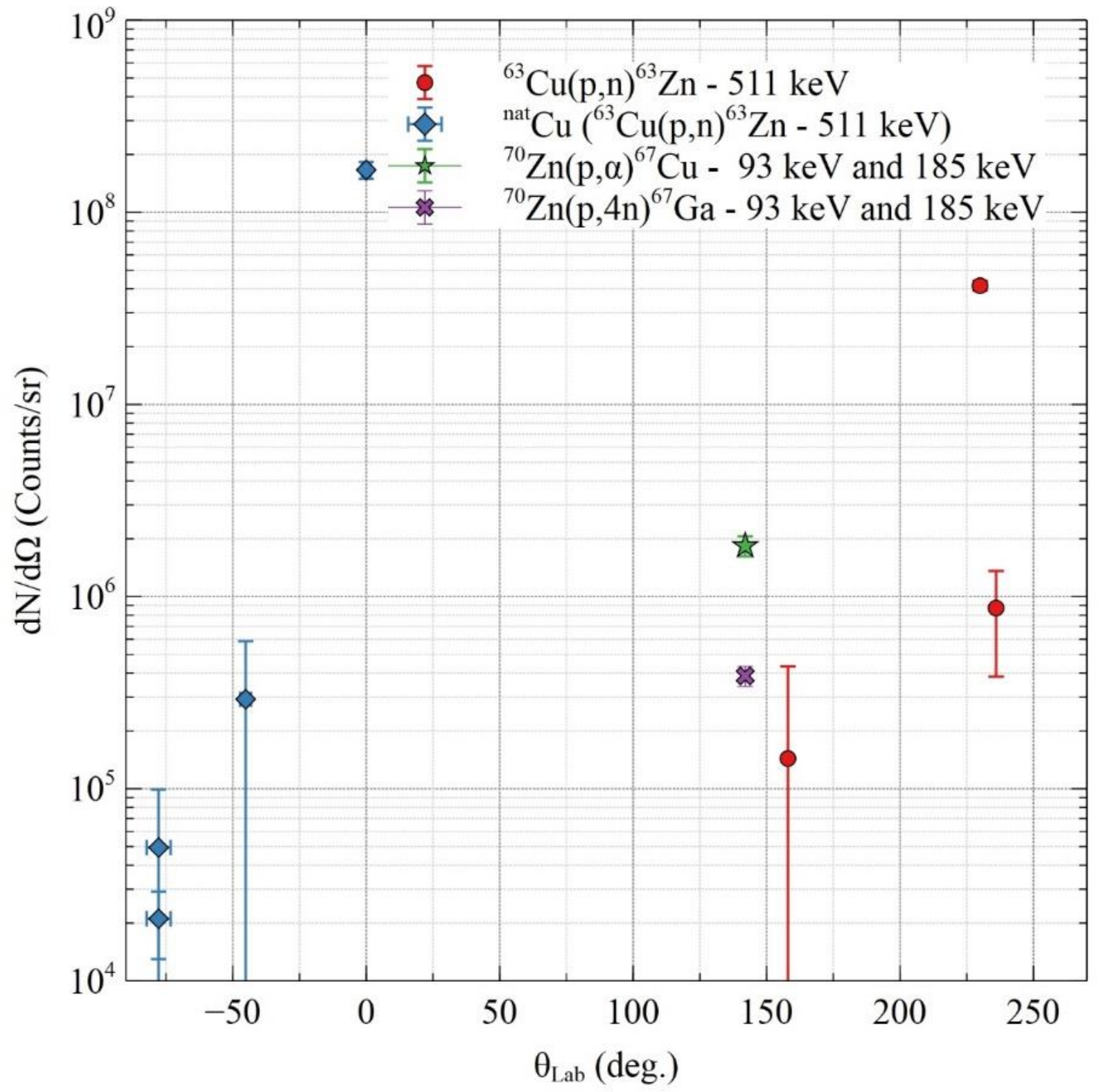

**Figure 7.** Summary of the yields at different angles obtained using the targets indicated in the inset. Angles between -90 and +90 degrees are in the forward direction, from 90 to 270 degrees are in the back direction of the Al target producing protons through the "TNSA mechanism", see the Appendix A.

We may notice that we have two results at -90 degrees using the $^{nat}Cu$ target. For these two separate series of shots, a $^{nat}B$ target was installed in front of the Al target in the catcher-pitcher configuration (CP) [8]. Since the B target is rather thick, we expect the produced $\alpha$-particles to remain trapped apart from those produced at the surface and emitted in the back direction [8]. Thus, we installed the $^{nat}Cu$ at -90 degrees to capture the $\alpha$-particles and activate it through (but not only) reactions #10,11 in Table I. As we see from Fig.7 and as expected, at such angles the proton distribution should be depleted thus a signal coming from the $\alpha$-particles maybe detectable. The reactions #5-7 in Table I may occur in such configuration. The experimental $\gamma$-spectra for such cases do not show any activations which can be attributed to the $\alpha$-particles. Even from this negative result we may learn something, i.e. we could increase the probability of $\alpha$-reactions locating the Cu target closer to the B target thus increasing the solid angle. A better option is to use the CP configuration but with a mixture of B and another element which can be unambiguously activated by the $\alpha$-particles. In such a case an experimental constraint on the $\alpha$-distribution may be obtained [20].

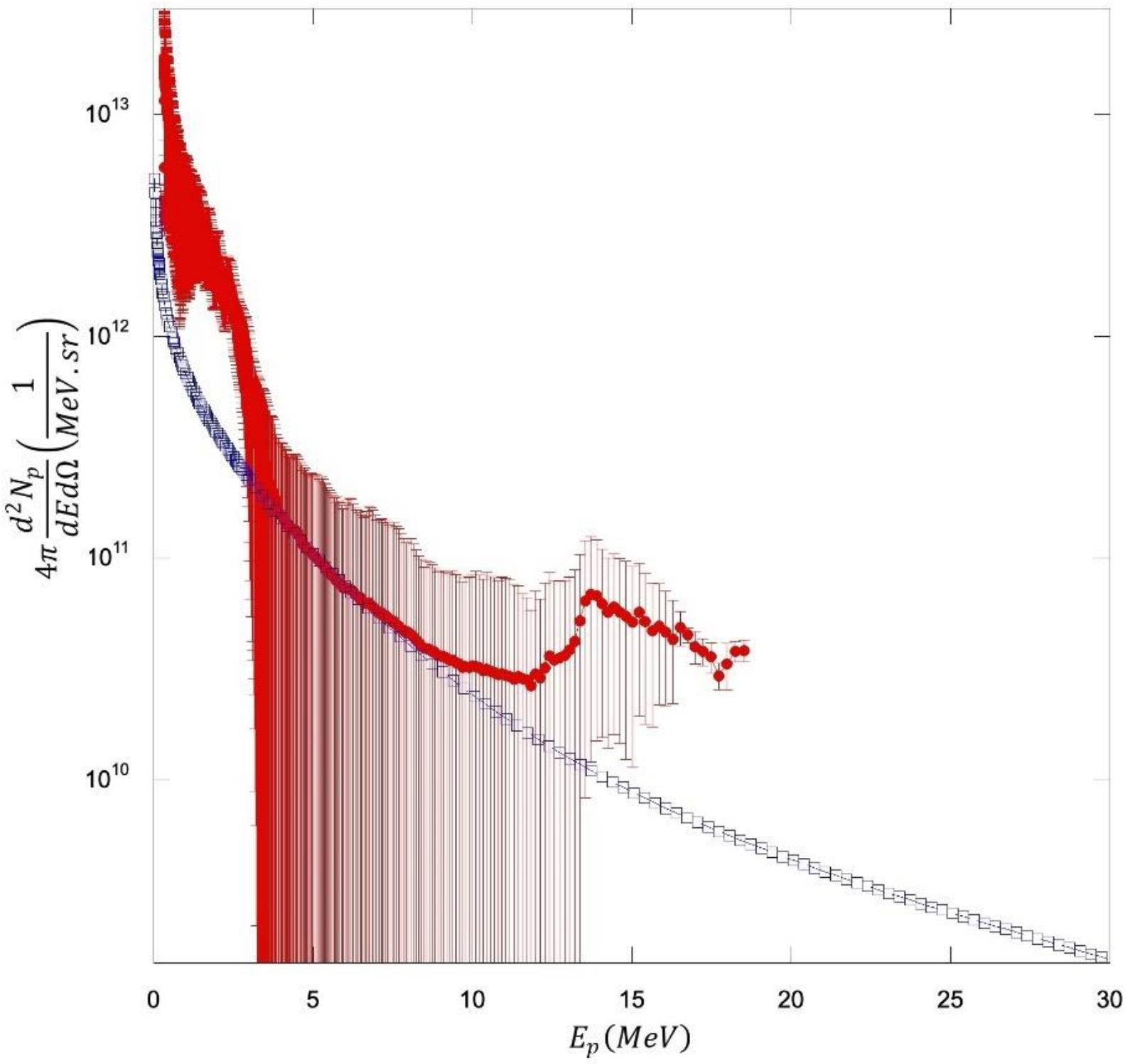

**Figure 8.** Proton distribution measured at zero degrees (full red circles) with the TP detector. The large error bars are fluctuations from shot to shot due to different experimental conditions [8]. The data are normalized to eq. (1) (open blue squares) with 4π b(0)=1.5±0.3 $sr^{-1}$

As it is shown in Fig.4, the $^{nat}Cu$ target was conveniently located in front of the Thomson Parabola (TP) spectrometer [8,9]. Proton energy distributions for each shot were recorded and their average value and fluctuations from shot to shot are displayed in Fig. 8, as compared to Fig 1. The TP is not calibrated thus we have normalized the data to our adopted proton distribution in the region 5-10 MeV where most of the $^{63}Zn$ yield is produced according to our calculations using eqs.(1-5). The parameter in eq.(1) 4π b(0)= 1.5±0.3 $sr^{-1}$, was fixed to reproduce the yield obtained with the $^{nat}Cu$ target at zero degrees. This may serve as a preliminary calibration, but the important point is that protons close to 20 MeV are observed and we cannot exclude that even higher energy protons are also produced, but their yield is below the sensitivity of the TP [8], see fig.8.

### 2.2 $^{70}Zn(p,\alpha)^{67}Cu$, $^{70}Zn(p,4n)^{67}Ga$

Reactions #3&4 in Table I, apart from being interesting for medical applications [20,22,24], may give some further constraint on the proton distribution (in the back direction) because of the different Q-value. In particular, to produce $^{67}Ga$, protons with energies higher than 28 MeV are needed and we did not observe any with the TP located in the front direction, see Fig.8. However, the proton distribution in the back direction can be quite different from that in the front one [20].

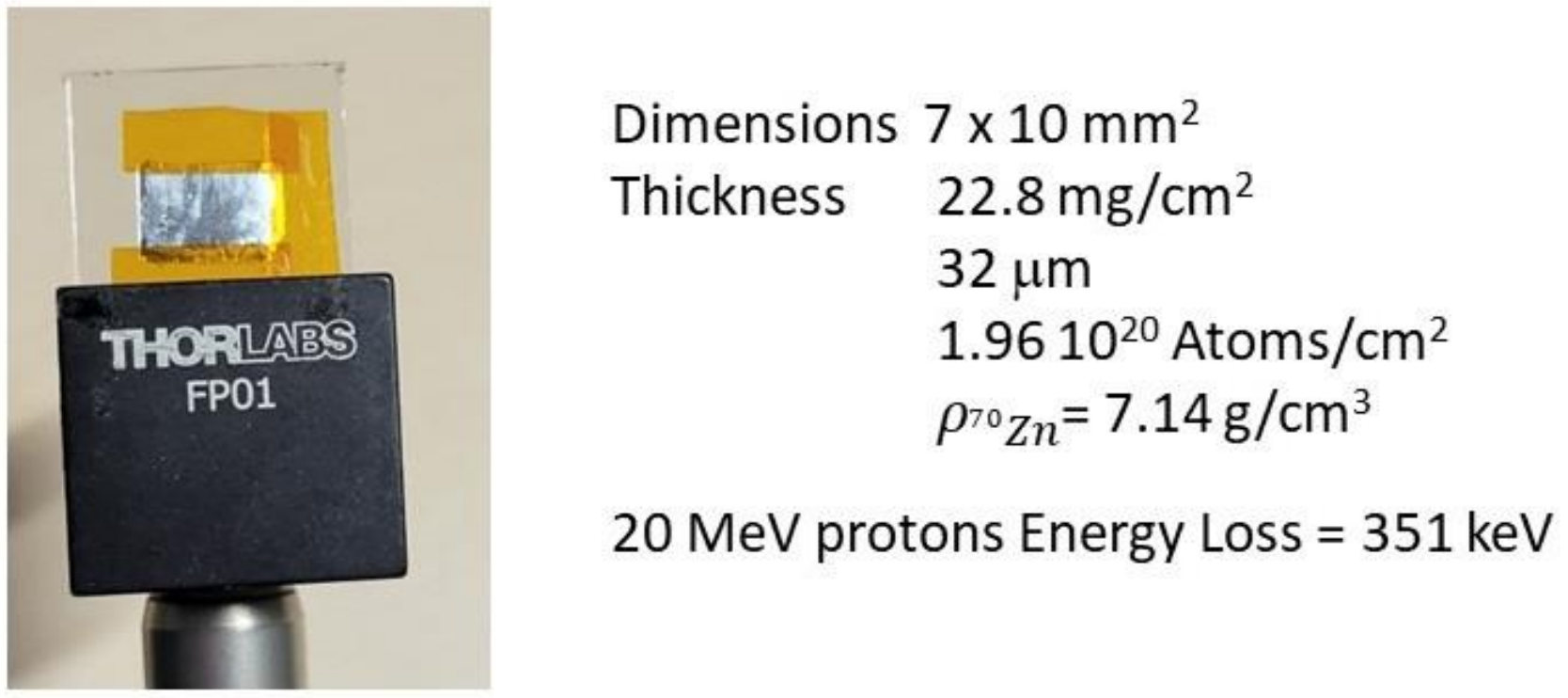

**Figure 9.** The $^{70}$Zn target mounted on the CR39 and installed at 130 degrees (back direction), see also Fig. 7.

In Fig. 9, we show the $^{70}$Zn target used in our experiment. Like the $^{63}$Cu case, the CR39 was added for support and to measure the proton distribution [8], however the support was taken off when measuring the γ yield using the HPGE detector. To estimate the yields for the reactions #3&4 in Table I, the cross sections are needed. The $^{67}$Cu case has been rather well studied and the cross section is well known while the data is scarce for the $^{67}$Ga case. As we discussed above, we parametrized the cross sections using eqs. (4) and (5). For the $^{67}$Ga case some data are available but at high proton energies [25]. We used four fitting parameters to reproduce the proton energy threshold value and the data at high energies.

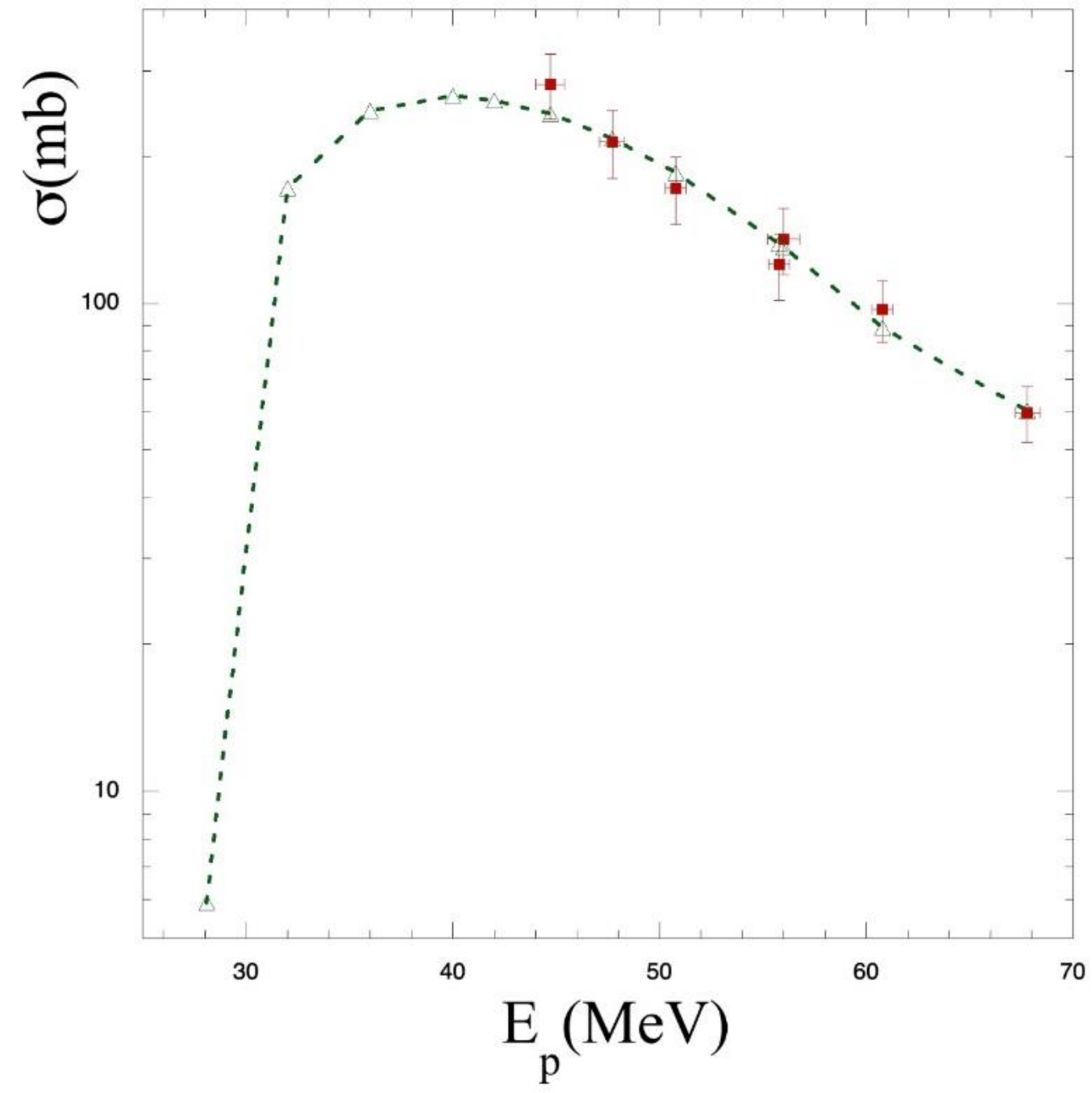

**Figure 10.** Cross section for the reaction $^{70}$Zn(p,4n)$^{67}$Ga as a function of proton energy. The data are taken from ref. [25], full squares. The parametrization using eqs. (3) and (4) is given by the open triangles joined by the dashed line.

In Fig. 10, we compare our parametrization to the data [25]. No data is available to our knowledge in the region near 30 MeV which gives the highest contribution to the yield reported in Table I.

Furthermore, the fact the yield of the two reactions is comparable (if we have proton energies higher than 30 MeV) is because the cross sections differ of more than one order of magnitude especially at high energies, see also Fig. 3 and Fig. 6 in ref. [25]. In particular, the reaction cross section to produce $^{67}$Cu reaches at most 20 mb while for $^{67}$Ga about 300 mb were measured [25].

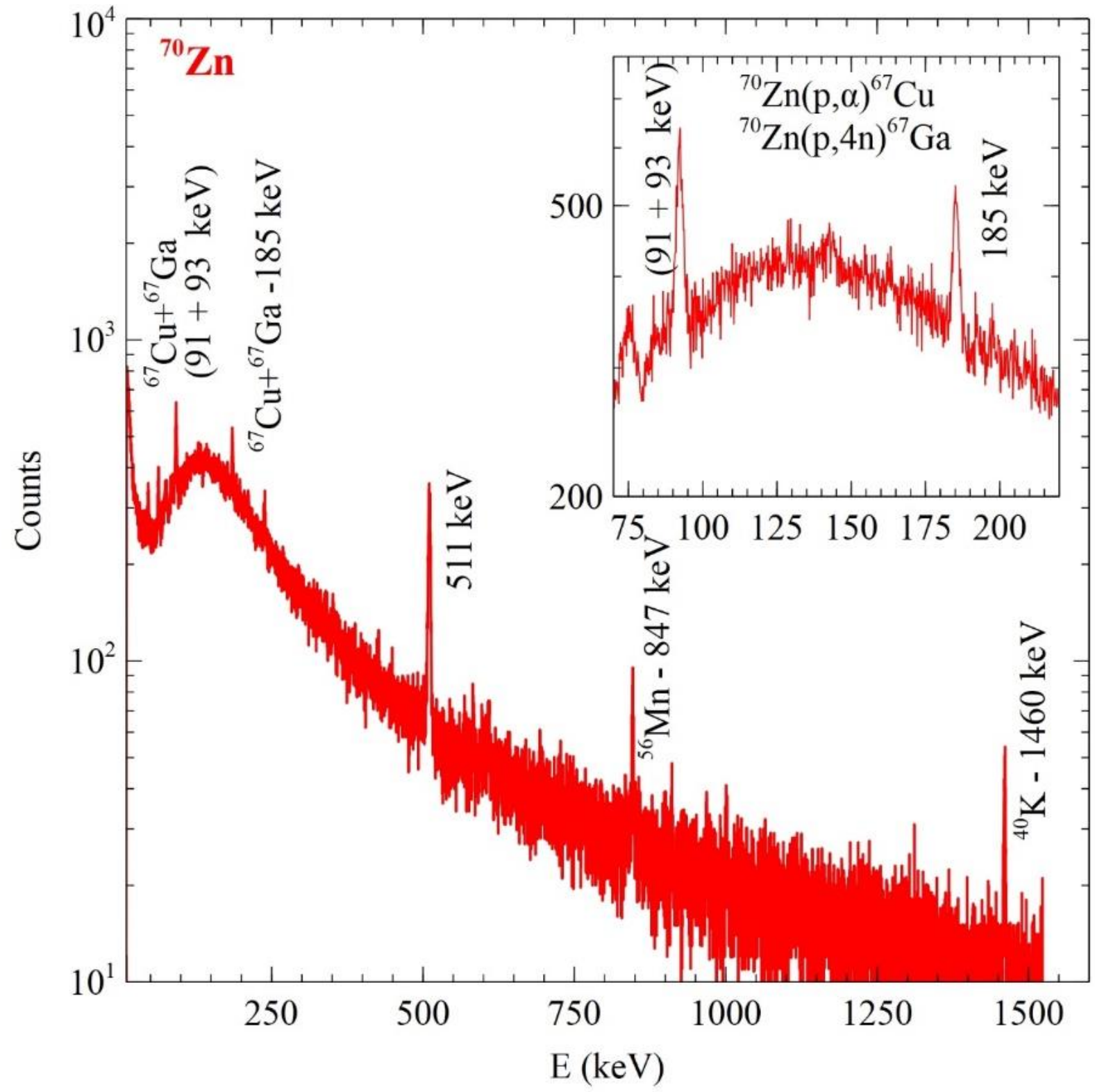

**Figure 11.** The β-delayed γ-ray spectrum for $^{70}$Zn target. The FEPs 91-93 keV (within the HPGE energy resolutions) and at 185 keV contain the contributions from the decay of $^{67}$Cu and $^{67}$Ga produced in the $^{70}$Zn(p,α)$^{67}$Cu and $^{70}$Zn(p,4n)$^{67}$Ga reactions. This spectrum was measured with the original acquisition system (see Appendix B) and it is not background subtract. The FEPs count rate for 511 keV and 1460 keV are 0.012 and 0.0014 counts/s compatible with the background measured with the new acquisition system. We have also identified a peak corresponding to $^{56}$Mn probably coming from activation of some impurities. Another possibility is given by the reaction $^{70}$Zn(p,$^{56}$Mn)$^{15}$C; Q = -15.2 MeV, $T_{1/2}$ = 2.6 h.

The two peaks corresponding to the decay of $^{67}$Ga and $^{67}$Cu at 91-93 keV and 185 keV are clearly visible in Fig. 11, see also the inset in the figure for the region of interest. Note that the lower energy peak (91+93 keV) is more intense than the one at 185 keV. The gamma intensities per 100 decays of the β-decay parent for $^{67}$Cu are Iγ(%) = 13.2 and 48.7 for the two energies [17], thus we would expect the 185 keV to be more intense. On the other hand, for $^{67}$Ga Iγ(%)=42.4 and 21.2 respectively [17], see Table I. Using the well-known intensities for the two nuclei, we can solve a simple set of two equations with two unknowns to obtain their respective yield. The obtained values are plotted in the Fig. 7 with star and cross symbols, they are much smaller than the $^{63}$Zn case as expected from Table I but higher than our estimates, maybe suggesting that the proton yield in the back direction is higher than in the front [19]. This is consistent with the high yield seen for the $^{63}$Cu case at 230 degrees.

## 2.3 $^{11}B(p,n)^{11}C$

In Table I, we have estimated the total yield for the reaction #6 in good agreement with ref. [5]. In the spirit of the catcher-pitcher configuration [8], we located a thick natural boron target at few cm from the "TNSA" Al target. The B target was slightly tilted with respect to the normal direction to 'focus' some $\alpha$-particles produced at the surface to -90 degrees where we installed the $^{nat}Cu$ target, another TP and CR39 [8], see the appendix A. Many shots were performed on each target and gamma decays from the $^{nat}B$ target were measured with the HPGE detector. The $^{11}C$ yields for different measurements are reported in Fig. 12 together with pictures of the B targets after irradiation. The slightly ellipsoidal proton spots on the targets can be clearly seen, and we can measure their sizes. For instance, the spot for target 5-18 Nov gives an area of the order of 1 cm$^2$. The distance from the Al target is well known and this gives the protons opening angle $\Delta\Omega = 0.59\pm0.18$ sr which we already used above. Since all the produced protons are collected by the $^{nat}B$, the measured $^{11}C$ rate is the total yield in the front direction. As we see from Fig. 12, the measured yield is very close to our estimate in Table I [5] and could be probably improved a little. Notice that for the first 2 cases in Fig. 12, the yield was a factor of 2 less than our estimate because a little space was left between two $^{nat}B$ to let protons get through and be measured with a TP and/or diamond detectors [8].

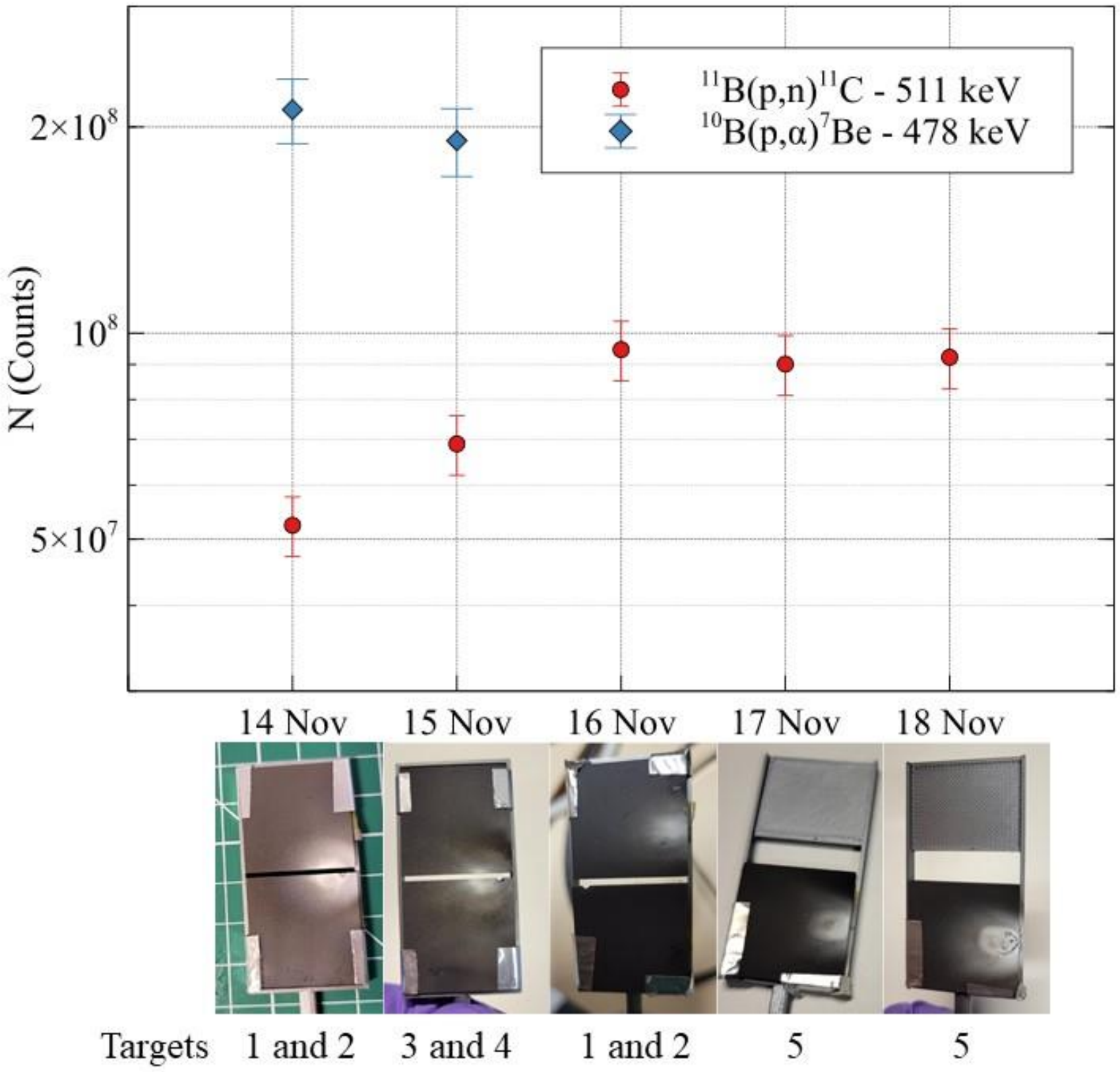

**Figure 12.** $^{11}C$ (full circles) and $^{7}Be$ (full diamonds) yields produced in TNSA protons colliding with $^{nat}B$ targets displayed in the figure at the end of each measurement. The proton spot size is clearly visible, and its opening angle can be estimated as discussed in the text. All results are corrected for the $^{11}B$-$^{10}B$ concentrations.

The used targets had a 99.6% concentration of $^{nat}B$ (80% $^{11}B$, 20% $^{10}B$) with remaining contaminants, mainly N, O and H. In refs. [20,26] the possibility was discussed that produced $\alpha$ particles may interact with the N impurities and produce $^{18}F$, i.e. ternary collisions [16,27]. It decays with 511 keV γ-rays just as $^{11}C$ but different lifetime. $^{13}N$ and $^{18}F$ can be produced by protons colliding with target 'impurities' ($^{14}N$ or $^{18}O$) [20]. Those elements have a very small concentration in the target (but much higher than the produced α through reactions #5,7 in Table I) and we expect a negligible contribution but in competition with ternary collisions.

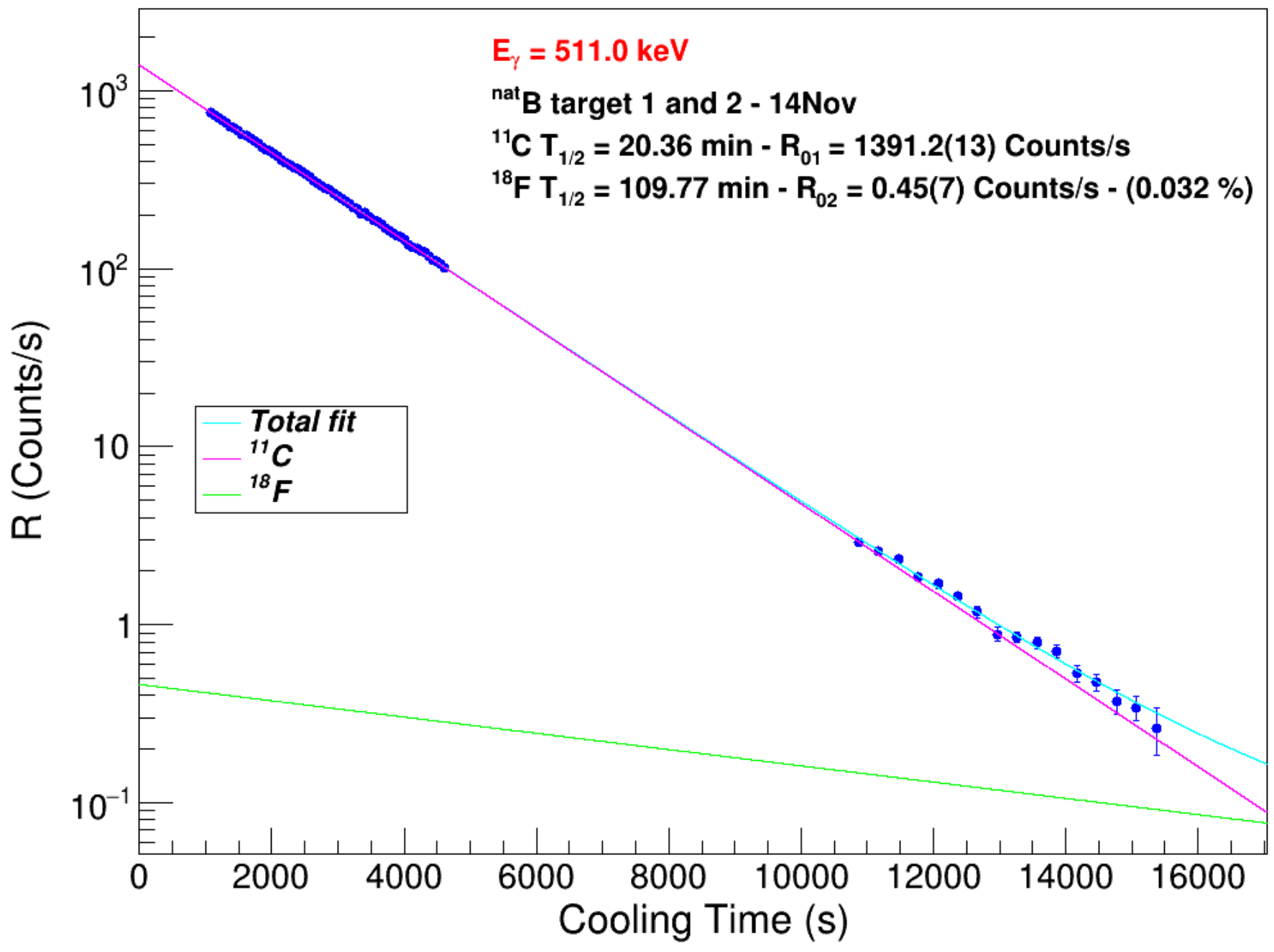

**Figure 13.** Counting rate for the 511 keV γ-rays emitted from the B target as a function of time. The fixed half-lives ($T_{1/2}$) are indicated in the figure. The rate can be fitted with just the $^{11}C$ decay. To put an upper limit for other ions, we have forced the fit using the half-lives of $^{11}C$ and $^{18}F$. The data support less than 0.032% $^{18}F$ produced for targets 1 and 2 in Fig. 12.

In Fig. 13, we plot the 511 keV γ counting rate for the B target in the pitcher catcher configuration. A fit to the experimental curve with free fitting half-lives and assuming three components gives 100% contribution from the decay of $^{11}C$. To obtain an upper limit to the production of the two other isotopes, we fixed their half-lives and rerun the fit. This gave a zero contribution from $^{13}N$, which is not surprising since its half-life is shorter than $^{11}C$, thus its contribution is relevant at short times when $^{11}C$ is dominant. $^{18}F$ has a longer half-life and it could be seen from the decay rate at longer times. In Fig. 13, we show the result of the fit fixing the half-lives of $^{11}C$ and $^{18}F$. This is an upper limit of 0.032% contribution of the $^{18}F$, like ref. [20], compatible with the low $^{18}O$ concentration, thus excluding any ternary collisions as expected [16,20,27]. This result was obtained in the situation indicated as target 1 and 2 in Fig. 12, i.e. a configuration where two B targets are slightly displaced. In this case, a certain number of protons propagated in vacuo and was detected by the TP at zero degrees. If we repeat the same analysis for $^{18}F$ but with the B target completely blocking the proton flow, cases 3-5 in Fig. 12, the production of $^{18}F$ increases to 0.15%. This difference may be simply understood by inspecting eq. 2 and realizing that all terms are the same for the two cases apart from the number of protons which must be decreased.

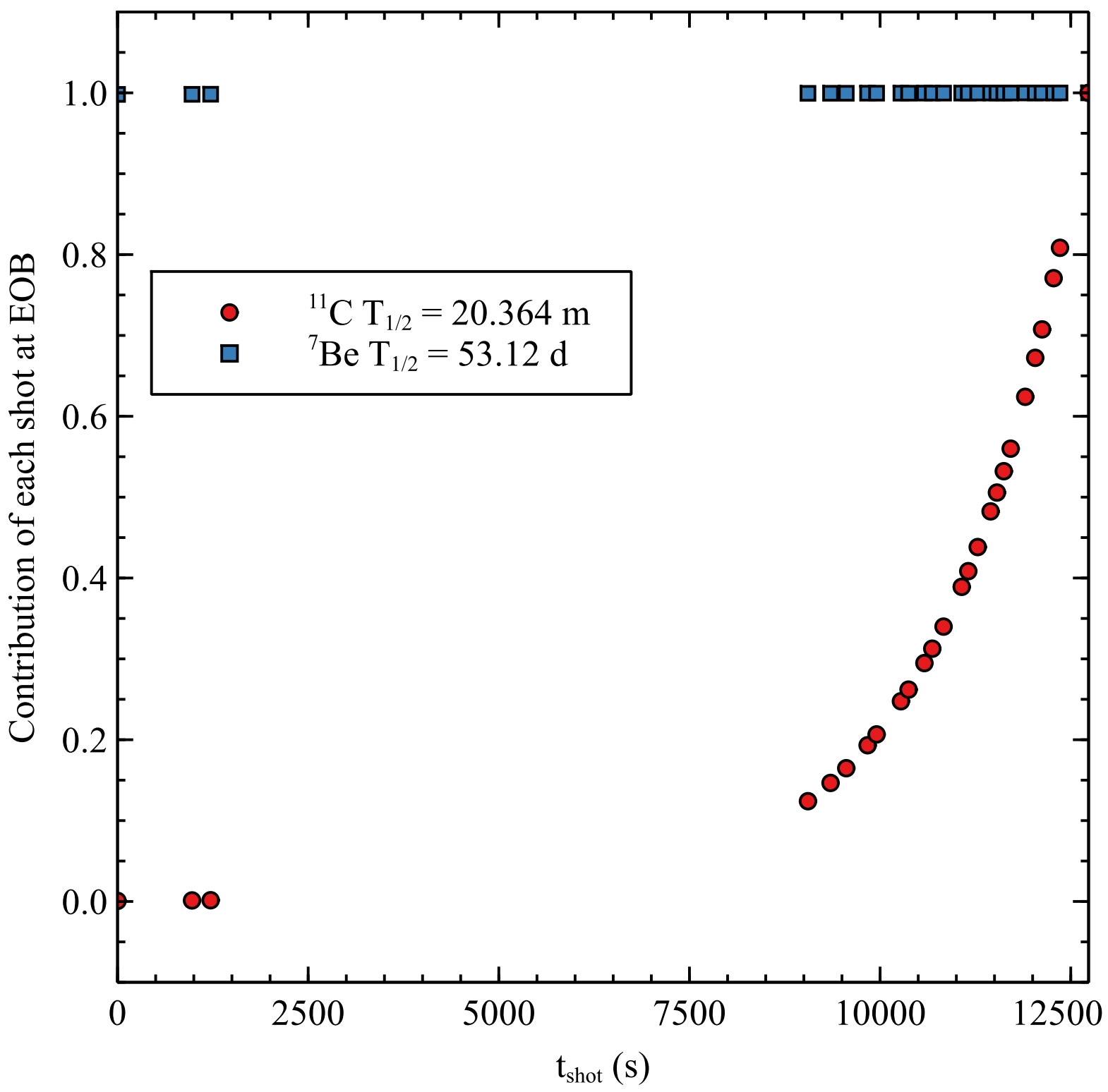

**Figure 14.** Probability contribution of each shot to the production of $^{11}C$ and $^{7}Be$ isotopes at End of Bombardment (EOB) as a function of shooting time starting from the first one ($t_{shot}$ = 0 s) to the last one when the chamber is opened, and the sample decays were measured using the HPGE detector. The well-known half-life ($T_{1/2}$) of both isotopes were used to produce this result.

### 2.4 $^{10}B(p,\alpha)^{7}Be$

In the $^{nat}B$ target we can also produce $^{7}Be$ from reaction #5 in Table I [15]. Its half-life is much longer than the previous case, which means that the shot contribution at different times has practically weight 1 as shown in Fig. 14 (compare to Fig.3).

The $\gamma$ spectrum for these reactions is dominated by $^{11}C$ at shorter times, while at longer times only $^{7}Be$ remains. Furthermore, the $\gamma$ energy decay of $^{7}Be$ (478 keV) is quite different from $^{11}C$ (511 keV) and can be quite easily distinguished in the plot, see Fig. 15.

The produced yield of $^{7}Be$ is displayed in Fig. 12 and it is in fair agreement to our estimate, reaction #5 in Table I, after correcting by the concentration. Since we only had one HPGE detector available, the two first cases only were measured for a long time to allow an unambiguous determination of $^{7}Be$.

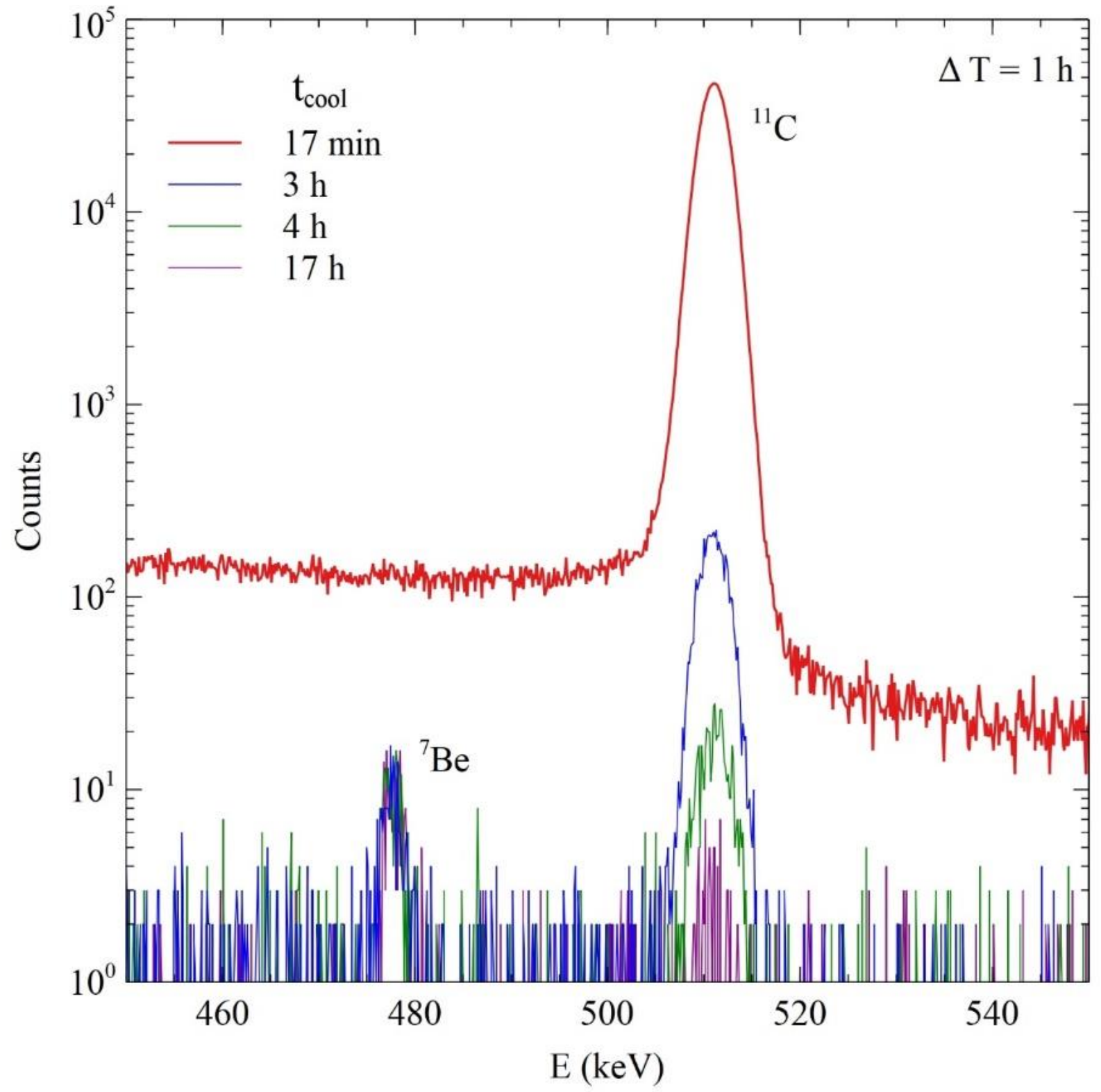

**Figure 15.** The region of interest for the background-subtracted β-delayed γ -ray spectrum for the $^{nat}B$ targets 1 and 2 exposed 14nov at four different cooling times ($t_{cool}$) for a measurement time of $\Delta T$ = 1h. The shortest cooling time is dominated by the $^{11}C$ decays while $^{7}Be$ becomes dominant at longer cooling times.

## 3 Conclusions

We have investigated nuclear reaction products using the PW laser at Vega III in Salamanca-Spain. The pitcher-catcher method was adopted with protons produced by an aluminum target and impinging on several different targets, both in the forward and the back direction, with respect to the laser one. We found the production of medical radioisotopes in agreement with expectations and the predictions of ref. [5], thus supporting its conclusions. Laser technologies are mature enough to compete to accelerators for the production. It may be a question of costs for the construction, spaces, maintenance etc.., to attest to their competitiveness but from our experience we predict that this may be a winning technology.

We have also estimated theoretically the production of different nuclei and made some assumptions for the cross sections in energy regions where they have not been measured. The "TNSA mechanism" should be studied more in detail and adjusted to the various physical scenarios that one would like to implement. In particular, the role of the electrons must be clarified and if possible, used to favor nuclear reactions in the plasma. This may be crucial if we would like to apply this method to neutronless reaction energy production. Our predictions in Table I are rather well confirmed by the data we obtained. The reaction $^{11}B(p,\alpha)^{8}Be$ will be discussed more in detail elsewhere [8]. The values obtained here are too small to have self-sustained reactions and we may need to compress the catcher [16,27].

## 4 Author Contributions

AB devised the scheme to measure radioisotopes and wrote the first draft of paper. MRDR performed the data analysis of the HPGE and ME the data analysis of the zero-degree TP. FaC was the spokesperson of the main experiment for α production. All authors contributed to the experiment proposal, preparation and on the final form of the manuscript.

## 5 Funding

This work was supported in part by the United States Department of Energy under Grant #DEFG02-93ER40773; the EUROfusion Consortium, funded by the European Union via the Euratom Research and Training Programme (Grant Agreement No 101052200 — EUROfusion); the COST Action CA21128- PROBONO "PROton Boron Nuclear fusion: from energy production to medical applicatiOns", supported by COST (European Cooperation in Science and Technology - www.cost.eu).

## 6 Acknowledgments

A particular thank you to all the personnel at the CLPU-Vega facility for their direct and indirect help. We thank Mr. A. Massara (LNS-INFN, Catania-Italy) for providing to specification the thin targets used in the experiment. We thank Dr. B. Roeder for the digital acquisition support.

## 8 Appendix A

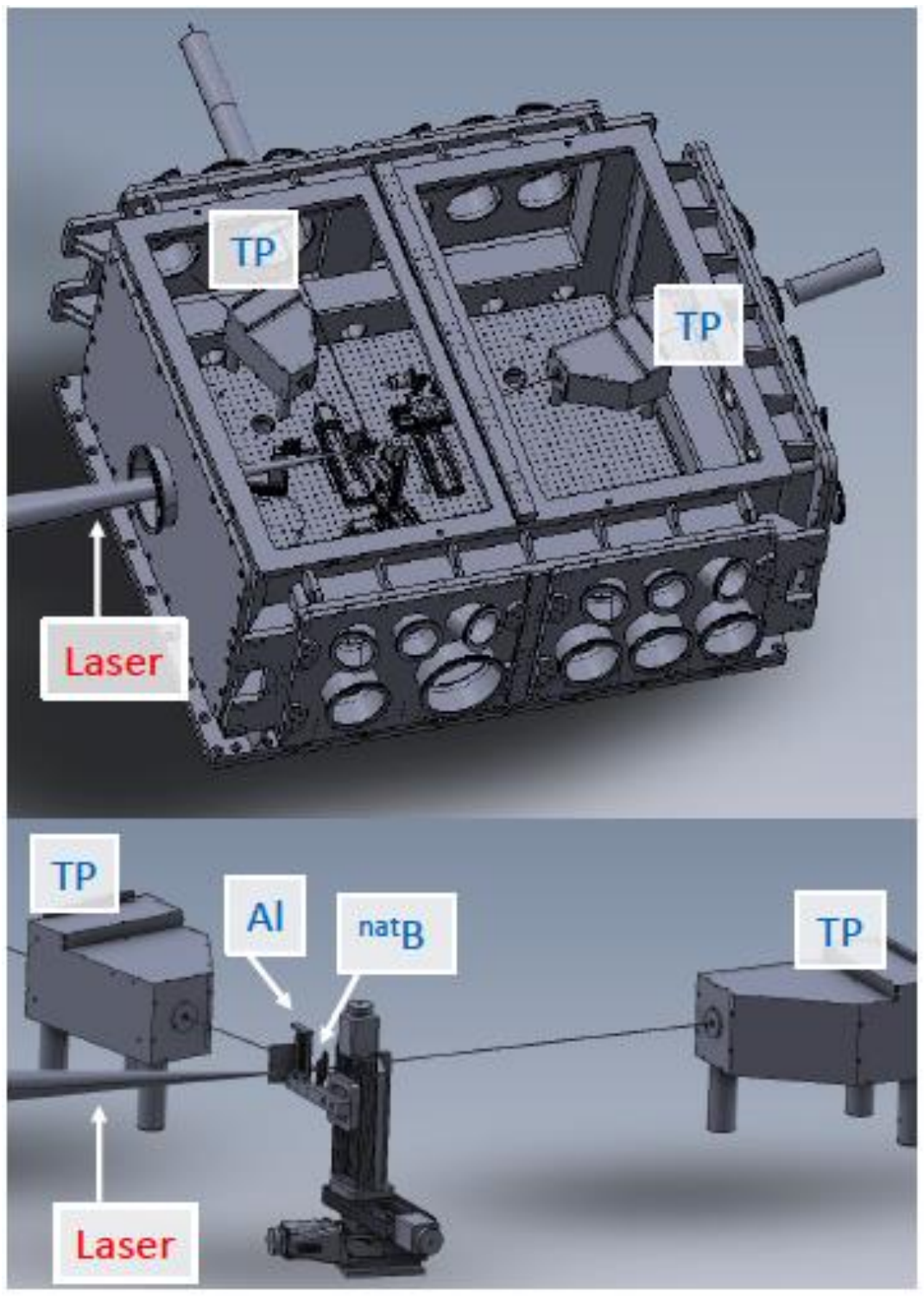

**Figure A1.** Schematic view of the scattering chamber. Top panel: the laser is impinging from the left on the Al target (to produce protons). The Al target is slightly tilted with respect to the laser direction to avoid dangerous reflections. The protons move in the direction perpendicular to the Al target and are detected by the forward TP which defines zero degrees, while the second TP is located at -90°. Bottom panel: the $^{nat}B$ is in front of the Al to collect the protons produced in the forward direction. In this Pitcher-Catcher configuration the TP at zero degrees is (partially) impeded.

### 8.1 Appendix B

#### Detector properties and calibration

A High Purity Germanium (HPGe) detector Canberra XtRa™ (coaxial germanium detector with a thin-window contact on the front surface), model GX3019[(1)], with a passive iron shield of 15 cm in all directions (see Fig. B1) was used for the γ - spectroscopic analysis. The original acquisition system is composed by a DSA-1000, a 16k Multichannel Analyzer (MCA) based on digital signal processing techniques, paired with GENIE 2000 (comprehensive environment for data acquisition, display, and analysis of γ -spectrometry) [(1)]. This system was used just for the measurement of $^{70}Zn$ target. Another updated acquisition system was used for all the other measurements using a CAEN DT5781 quadruple independent 16k digital MCA[(2)] and the CoMPASS[(2)], a multiparametric data acquisition software for physics applications. The main advantage of this set up is the capability to have a timestamp for each event, besides the use of an updated computer system.

The HPGe detector dead time was 1-10% and the energy resolution 1.0–2.5 keV (FWHM). The distances of detector cap to the sources were in the range of 20 and 120 mm. The 20 mm distance was only used for sources with low activities. The γ-ray spectra were measured and analyzed up to 1600 keV. Measurements of the $^{22}$Na, $^{155}$Eu and $^{137}$Cs sources were performed at the detector cap to the source distances in different days. These measurements were used in the detector energy and efficiency calibration. The absolute photopeak efficiencies for the different source-detector distances and source geometries were determined using the results of the experimental efficiencies from the sources measurements and the LabSOCS (Laboratory Sourceless Calibration Software) mathematical efficiency calibration software[(1)]. For instance, efficiency curves generated by LabSOCS considering plaques sources of 10 x 10 mm$^2$, 25 x 50 mm$^2$ and 100 x 100 mm$^2$ are presented in Figure B2 for illustration. The room background was measured during the experiment and subtracted in the analysis.

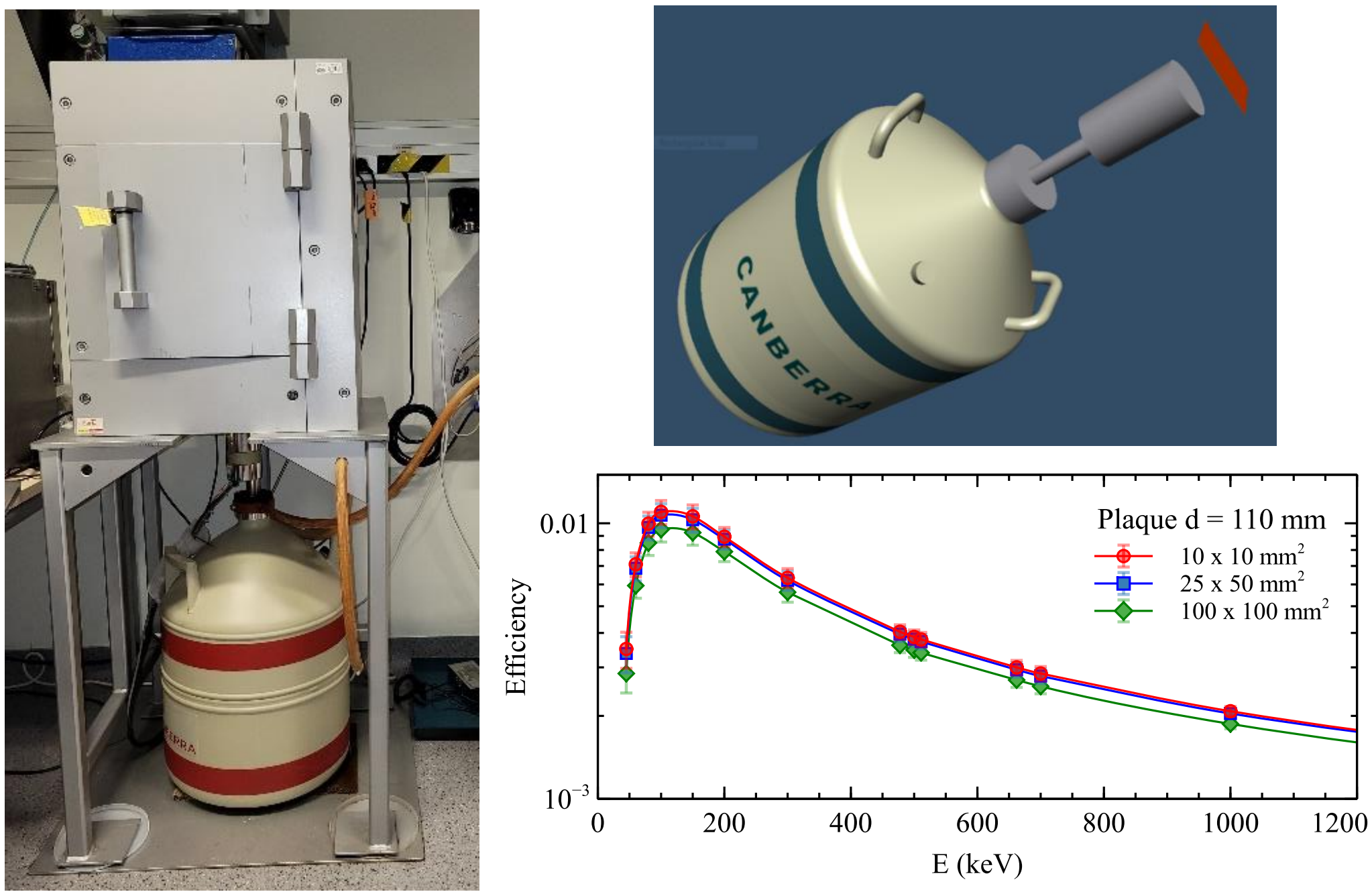

**Figure B1.** On the right panel there is a photo of the HPGe detector and the passive iron shielding. On the right panel there are the drawing generated by LabSOCS for a plaque source of 100 x 100 mm$^2$ at 110 mm from the detector cap and the respective efficiency curve. The efficiency curves for source plaques of 10 x 10 mm$^2$ and 25 x 50 mm$^2$ are presented for comparison.

### Spectrum and time evolution analysis

The data consists of the energy and timestamp for each event as described in the previous section. One example can be observed in Fig. B3 in the energy versus timestamp histogram. Depending on the half-life of each radionuclide of interest, spectra with different intervals of time (ΔT) are generated. In Fig. B3 are presented the spectra generated for ΔT = 60s to analyze the time evolution of $^{11}$C ($T_{1/2}$ = 20.36 min). The room background spectrum is normalized to ΔT and subtracted from each generated spectrum. The full-energy-peaks (FEPs) net area were obtained through a gaussian and background fit considering the Compton scattering background. In cases where the statistics is too low and the

(1) www.mirion.com
(2) www.caen.it

gaussian fit is not adequate to determine the area, the net area is obtained with the integral of the raw spectra in a fix width and subtracted by the integral of the room background spectrum normalized to ΔT in the same width.

The next step is the fit of the FEP rate as a function of time to identify the contribution of possible radionuclides that had enough different half-lives to be separated in the fit, see Fig. 6 and 13. The full spectrum analysis is also taken into consideration to determine any possible coproduced radionuclide with similar half-lives. Once the identification is done the number of radionuclides measured is determined using the respective FEP rate:

$$N_{exp} = \frac{R}{\lambda \varepsilon_{HPGe} I_{\gamma}} \qquad \text{(B1)}$$

Where R the photo-peak rate at energy $E_{\gamma}$, $\varepsilon_{HPGe}$ the detector efficiency of the HPGe detector at the γ -energy considered, $I_{\gamma}$ the intensity of the γ-line of interest, λ the decay constant of the radionuclide. Considering the production and decay in a multi-shot accumulation regime the number of radionuclei produced is given by:

$$N(t) = \sum_{i=0}^{n} N_{0i} e^{-\lambda(t - t_{shot\ i})} \approx N_0 \sum_{i=0}^{n} e^{-\lambda(t - t_{shot\ i})} \qquad \text{(B2)}$$

for $t > t_{shot\ n}$ , where t = 0s is the time of the first shot and $t_{shot\ i}$ is the time of the subsequent shots, n is the number of shots, $N_{0i}$ is the number of particles produced at shot $i$. If an average number of particles produced at each shot as $N_0$ is assumed, the approximation on right side of eq. B2 can be used to fit the $N_{exp}$ data and obtain the value of $N_0$, as seen on Fig. B4.

The corrections due to backscattering and coincidence summing peaks for $^{11}C$, 0.17-0.35%, γ self-absorption up to 4%, the respective targets isotopic enrichment, were also included in the results. The uncertainties are propagation of the full-energy-peaks (FEPs) net area and the detector efficiency uncertainties. The detection lower limit identification was 0.015(3) particles/s.

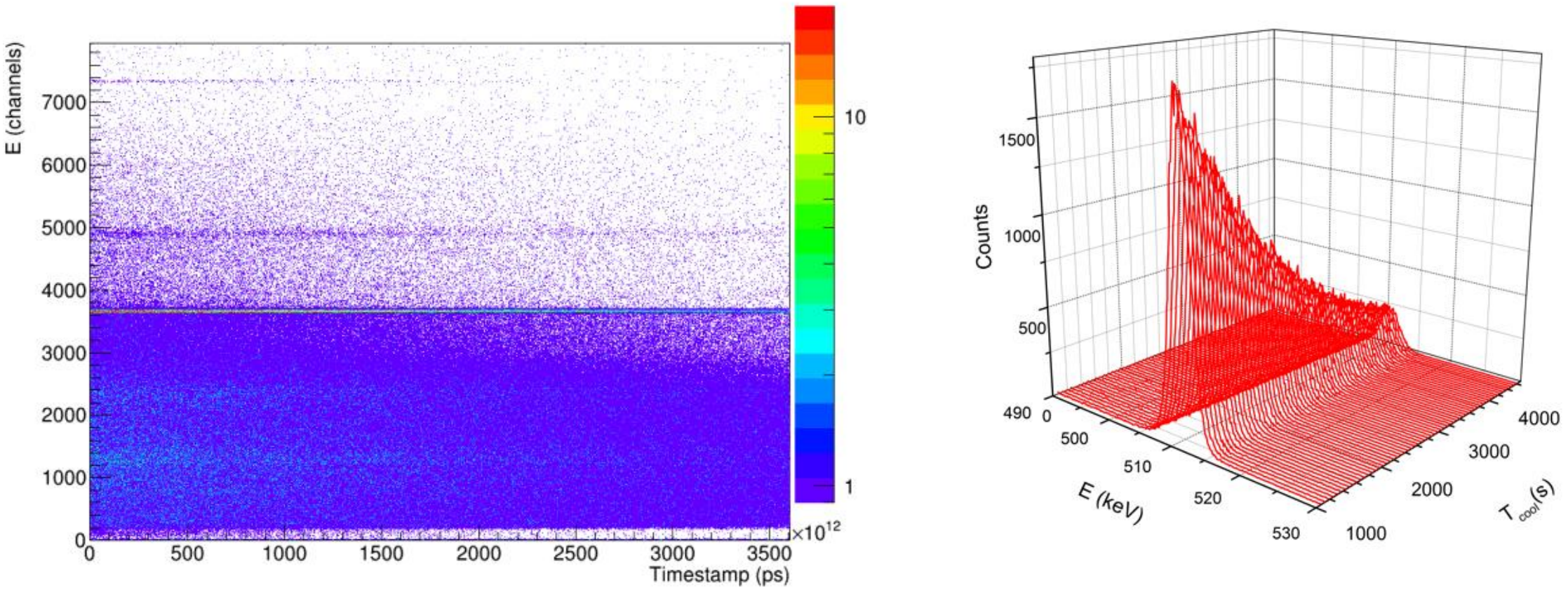

**Figure B2.** Energy versus timestamp raw data and the respective spectra generated using 60 s time intervals for $^{11}C$ analysis. The line at E(channel) = 3671 corresponds to the FEP at energy $E_{\gamma}$ = 511 keV.

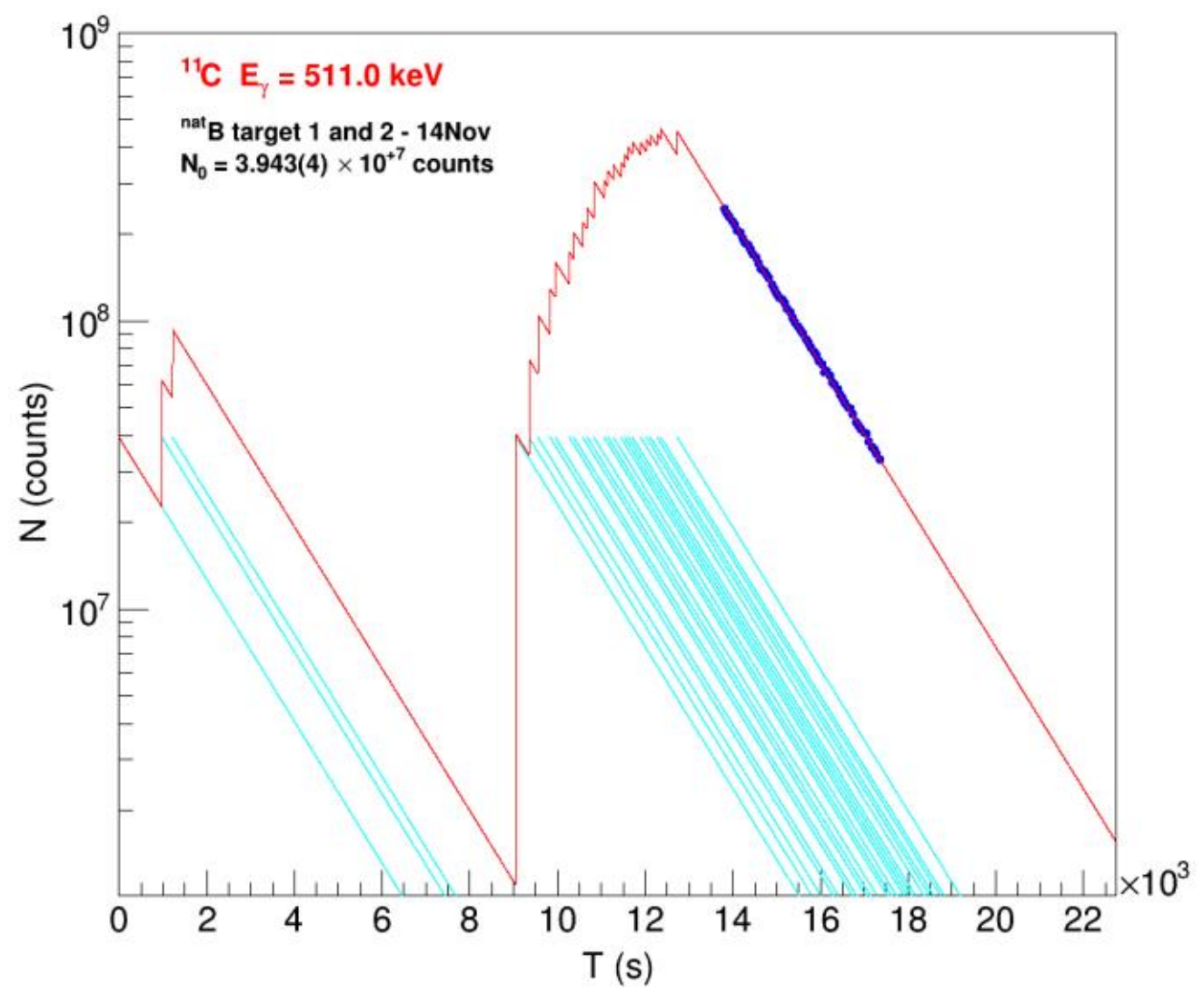

**Figure B3.** Number of radionuclides produced as a function of time. The first laser shot is at T = 0s. The blue points are the number of radionuclides obtained with the γ - spectroscopic analysis (eq. B1). The solid red line represents the eq. B2 fitted to experimental data. Jumps in the plot correspond to the time when a laser shot occurred. The cyan lines are the contribution for each shot.